\newcommand{\fig}[1]{\mbox{Figure\hspace{0.2em}\ref{#1}}}
\newcommand{\eq}[1]{\mbox{Equation\hspace{0.2em}\ref{#1}}}

\newcommand{\sect}[1]{\mbox{\S\ref{#1}}}

\newcommand{\ts}{\theta_\star}

\documentclass[preprint]{aastex}

\usepackage{graphicx}
\usepackage{natbib}
\usepackage{amssymb}
\usepackage{longtable}
\usepackage[nooneline]{caption2}

\begin{document} 
\title{Searching for sub-kilometer TNOs using  Pan-STARRS video mode  lightcurves: 
Preliminary study and evaluation using engineering data}

\author{
J.-H.~Wang\altaffilmark{1,2},
P. Protopapas\altaffilmark{3,4},
W.-P.~Chen\altaffilmark{2}, 
C.~R.~Alcock\altaffilmark{3},
W.~S.~Burgett\altaffilmark{6},
T. Dombeck\altaffilmark{5}
J.~S.~Morgan\altaffilmark{6}, 
P.~A.~Price\altaffilmark{6},
 and
 J.~L.~Tonry\altaffilmark{6}
}
\altaffiltext{}{email: jhwang@asiaa.sinica.edu.tw, pprotopapas@cfa.harvard.edu}
\altaffiltext{1}{Institute of Astronomy and Astrophysics, Academia Sinica.
  P.O. Box 23-141, Taipei 106, Taiwan}
\altaffiltext{2}{Institute of Astronomy, National Central University, No. 300,
  Jhongda Rd, Jhongli City, Taoyuan County 320, Taiwan}
\altaffiltext{3}{Harvard-Smithsonian Center for Astrophysics, 60 Garden Street,
  Cambridge, MA 02138}
  \altaffiltext{4}{Initiative in Innovative Computing, School of Engineering and Applied Sciences, 29 Oxford Street,
  Cambridge, MA 02138}
\altaffiltext{5}{Physics Department, University of
Hawaii, 2680 Woodlawn Drive, Honolulu, HI, 96822} 
\altaffiltext{6}{Institute for Astronomy, University of
Hawaii, 2680 Woodlawn Drive, Honolulu, HI, 96822} 
 

\begin{abstract}
We present a pre-survey study of using Pan-STARRS  high sampling rate video 
mode guide star images to search for TNOs.  
Guide stars are primarily used by Pan-STARRS to compensate for
image motion and hence improve the  point spread function. 
With suitable selection of the guide stars within the Pan-STARRS 7 $\rm deg^{2}$ field of view, 
the lightcurves of these guide stars can also be used to search for occultations by TNOs. 
The best target stars for this purpose are stars with high signal-to-noise ratio (SNR) and small angular size.
In order to do this, we compiled a catalog using the SNR calculated  from stars with $m_{\rm V} <13$  mag in the Tycho2 catalog then cross matched these stars with 
the 2MASS catalog 
 and estimated their  angular sizes from $(V-K)$ color. 
We  also outlined a new detection method based on matched filter
that is optimized to search for  diffraction patterns in the lightcurves due to occultation by sub-kilometer TNOs. A  detection threshold  is set to compromise between real detections and false positives. 
Depending on the theoretical size distribution model used, we expect to find up to \emph{a hundred events} during the  three-year life time of the Pan-STARRS-1 project. The high sampling (30~Hz) of the project facilitates detections of small objects (as small as 400~m), which
are numerous according to  power law size distribution,  and thus allows us to verify various models and further constrain our understanding of the structure in the outer reach of the Solar System. We have tested the detection algorithm and the pipeline on a set of  \emph {engineering} data (taken at 10~Hz in stead of 30~Hz). No events were found within the engineering data, which is consistent with the small size of the data set and the theoretical models. Meanwhile, with a total of $\sim 22$ star-hours video mode data ($|\beta| < 10^{\circ}$), we are able to set an upper limit of  $N(>0.5 \rm ~km)\sim 2.47\times10^{10}$ deg$^{-2}$ at 95\% confidence limit.
\end{abstract}

\maketitle 
\section{Introduction}
\label{sec:intro}
Since the first Kuiper Belt object (not including Pluto)  was discovered in 1992  \citep{jewitt1993}, more than 1000  trans-Neptunian objects  (TNOs) have been 
found\footnote{See http://www.cfa.harvard.edu/iau/lists/TNOs.html for a list of these objects.}.  
A picture is emerging of the principal characteristics of the TNOs that is rich in information but leaves many unanswered questions. For example, a number of dynamical groups have been identified: classical, resonant,  scattered disk \citep{Luu1997}, and more recently the extended scattered disk \citep{gladman2002}. What  is the number and size distribution of the smaller objects of these dynamical groups? Is there an extension of the Kuiper Belt beyond 50 AU comprising bodies too small to have been detected in direct surveys?  Meanwhile, 
could a possible close stellar encounter in the early history of the Solar System \citep{Allen2001, Trujillo2001, bernstein2004, Fuentes2008} be responsible for the mass deficit and the depletion of larger objects (D $>$ 150~km) beyond 50~AU?

The size distribution of these small TNOs provides important clues to the dynamical evolution of the early Solar System. Kilometer-size TNOs with 4\% albedo are expected to have $M_{\rm R} \gtrsim30$, which is still way below the detection limit of the largest ground-based telescopes. Yet the vast majority of the TNO population is beyond the limit of direct observation --for example, the  Keck pencil beam survey \citep{Chiang1999} has a detection limit  of $M_{\rm R} \approx 27.5$.

A number of authors \citep{Bailey1976, Roques2000} have suggested the possibility of detecting TNOs by stellar occultation. An occultation manifests  itself as the shadow created by 
a TNO occulting a background star sweeping across the Earth, causing flux reduction in the  lightcurve. Hence, as opposed to direct observation, stellar occultation by TNOs provides a unique way of detecting kilometer to sub-kilometer size objects in the foreseeable future. Recently a few groups have attempted to search for occultations by TNOs \citep{Roques2006,bickerton2008,Lehner2009,kiwi2008,Bianco2009,bickerton2009}, but no real 
detection has been confirmed yet.

Projects committed to search for TNOs by occultation either do not have adequate time resolution, observe too few stars with good SNR or do not have enough observing time. These projects do a blind search for occultations, and these events are rare due to the reasons mentioned above.  Therefore, 
the key to a successful detection is to have  high sampling ($\ge$~20~Hz), high signal to noise ratio of background stars  (SNR $\ge$ 80),  small stellar angular sizes ($\leq$ 0.1~mas) and, most importantly, many star-hours of observations ($>$ a few hundred thousands). 

The Panoramic Survey Telescope and Rapid Response System (Pan-STARRS) is a project consisting of 4 telescopes that  can cover over  6000 $\rm deg^{2}$ per night or scan the whole visible sky from Hawaii   ($3\pi$) in a week to a detecting limit $\sim 24^{\rm th}$ magnitude. Pan-STARRS prototype telescope (PS1) will monitor up to 60 guide stars in a very high sampling rate video mode. 
We have compiled a list which allows us to select guide stars from anywhere in the field; guide stars 
could be in any of the 64$\times$64 OTCCDs (see $\S$\ref{sec:ps1}) and read at $\sim30$~Hz. 
Each field will have multiple choices of guide stars that were ranked based on the predicted event rates, which depend on their SNRs and angular sizes.  
By properly selecting guide stars,  the Pan-STARRS video mode images  would be ideal for searching TNOs. 

In the next section, we briefly describe the Pan-STARRS system. In $\S$\ref{sec:diff} and $\S$\ref{sec:det} we discuss the diffraction profiles and present our detection algorithm. In $\S$ \ref{sec:angular} we describe the prediction of stellar angular size using $(V-K)$ color. In $\S$\ref{sec:rate} we estimate the number of expected events based on different number density estimations, sampling rates and stellar angular sizes.  The methodology and  compilation of  the guide star list is in  $\S$\ref{sec:catalog}.  In $\S$\ref{sec:engdata}, we show and describe the quality of the engineering data obtained in fall 2008. Conclusions are in the final section.

\section{Pan-STARRS-1 system}
\label{sec:ps1}

The full version of Pan-STARRS will contain four identical telescopes and camera systems, which is also known as Pan-STARRS-4 \citep{kaiser2002}. PS1 is a single telescope and camera system built to test all requirements that are needed in the full version project.  PS1 telescope has an aperture of 1.8~m in diameter with 3-element wide field corrector that delivers a 7 deg$^{2}$ flat field in the focal plane. Its 8~m focal length gives a plate scale of 38.5 $\mu$m/arcsecond rendering a pixel resolution of 0.258 arcsecond/pixel. The telescope can monitor up to 240 guide stars at  $\sim30$~Hz (in reality, probably about 60 guide stars at 30Hz with high SNR). High frequency image motion can be removed by on-chip fast guiding, where localized image motion can be calculated from a collection of nearby guide stars. Hence different image motions across the field of view can be corrected. This can only be achieved with the innovative design of an orthogonal transfer charge coupled device (OTCCD). The OTCCD allows on-chip charge shifting, in both rows and columns, which can compensate for  the image motion in the focal plane \citep{1997PASP..109.1154T}. The PS1 Giga-Pixel Camera (GPC1) contains an 8x8 array, minus the 4
corners, of OTCCDs.  For reasons of economy of manufacture, facilitating
fast readout and other practical concerns, the OTCCDs each consist of an
8x8 Orthogonal Transfer Array (OTA) of independently-addressable cells
of size 590x598 pixels.  Hence GPC1 comprises approximately 1.4 billion
pixels. Typically, each OTA will have a single cell devoted to a guide star
reading at high frequency.  Of the 590x598 pixels of this cell, only a
small fraction around the guide star will be read, thus allowing a fast
read out with relatively low read noise.

\begin{table}[h]
\begin{center}
\caption{Filter bandpass, $m_{1}$ (the AB magnitude that produces $1\rm e^{-}$/sec/pixel) and  average sky brightness $\mu$ in magnitudes per arcsecond at Haleakala  for PS1 filters.}
\vspace{0.2cm}
\begin{tabular}{cccc}
\hline
Filter & Bandpass (nm) & $m_{1}$ mag & $\mu$ mag/$\rm arcsec^{2}$\\
\hline
g & 405-550 & 24.90 & 21.90 \\
r & 552-689 & 25.15 & 20.85 \\
i & 691-815 & 25.00 & 20.15 \\
z & 815-915 & 24.63 & 19.26 \\
y & 967-1024 & 23.03 & 17.98 \\
\hline
\label{tab:filter}
\end{tabular}
\end{center}
\end{table}

PS1 has four filters closely resembling SDSS filters ($g,r,i,z$), and a y-band filter. Their passbands, corresponding $m_{1}$ magnitudes and sky brightness are listed in Table \ref{tab:filter}. The major sources of noise are: shot noise, background noise from the sky, read noise, radio frequency interference and dark current noise.  The SNR for a given star can be estimated by\footnote{We took the numbers from PS1 document  PSDC-230-002-00 (Mission Concept Definition http://pan-starrs.ifa.hawaii.edu/project/PSDC/PSDC-2xx/) for the characteristic noise for all possible sources.},

\begin{equation}
\label{equ:snr}
S/N=S/\sqrt{\sigma_{\rm P}^{2}+\sigma_{\rm S}^{2}+\sigma_{\rm RN}^{2}+\sigma_{\rm RFI}^{2}+\sigma_{\rm T}^{2}+\sigma_{\rm d}^{2}}
\end{equation}
where the various sources of noise are
\begin{itemize}
		\item Poisson noise, $\sigma_{\rm P}^{2}=0.5\times10^{-0.4(m-m_{1})} \times t$, where $t$ is the exposure time and  $m$ is the apparent magnitude of 
		the target star.
		\item (sky) background noise,
		$\sigma_{\rm S}^{2}=\frac{\pi}{4}\omega^{2}\times10^{-0.4(\mu-m_{1})} \times t$, where $\omega$ 
		is the FWHM of the PSF in arcseconds, and $\mu$ is sky brightness in magnitude per square
		arcsecond;
		\item read noise, 
		$\sigma_{\rm RN}^{2}=\frac{\pi}{4}\omega^{2}\times A \times N_{\rm read}^{2}$, 
		where $A=3.846~\rm 
		pixel/\rm arcsec^{2}$ for 0.26 arcsecond pixels and  $N_{\rm read}$ is the read noise of the 
		detector.
		\item noise from radio frequency interference, $\sigma_{\rm RFI}^{2}=\frac{\pi}{4}\omega^{2}\times A \times N_{\rm RFI}^{2}$, where $N_{\rm RFI}^{2}$ is radio frequency interference  noise.
		\item noise from dark current, $\sigma_{\rm d}^{2}=\frac{\pi}{4}\omega^{2}\times A\times d \times t$, with $d$ counts/sec
\end{itemize}

\begin{figure}[]
\begin{center}
\includegraphics[width=90mm]{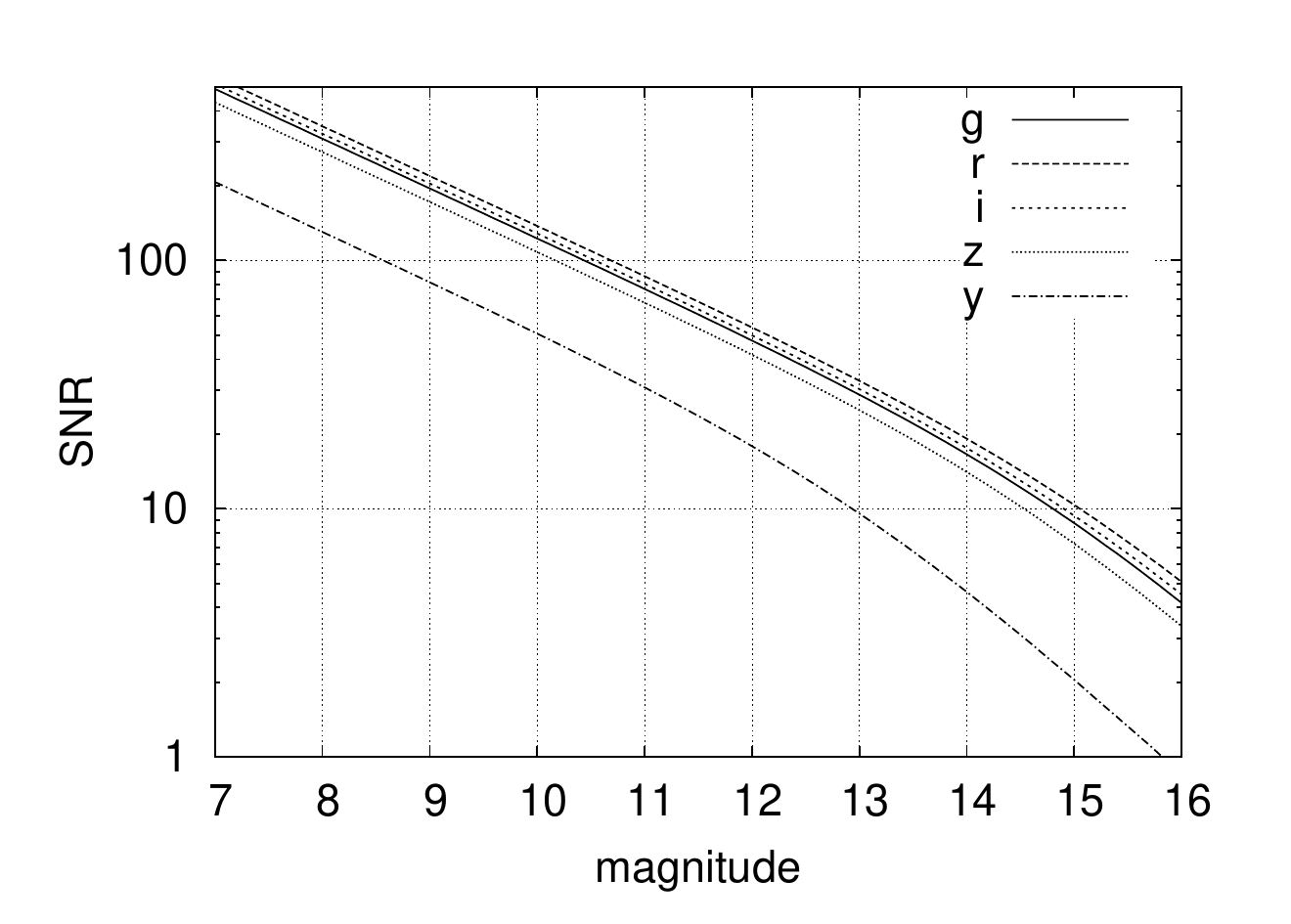} 
\caption{SNR plot for PS1 filters with 27$\rm e^{-}$ read out noise, 1~arcsec FWHM, $d=5$~counts/sec and 30~Hz sampling. Sky brightness and $m_1$ values for each filter are taken from  Table \ref{tab:filter}. Note that, except for the y filter, stars with $m_{\rm V} < 11$ mag can achieve SNR $\ge 80$ with 30~Hz sampling.}
\label{fig:snr}
\end{center}
\end{figure}

Based on the characteristic noise mentioned above, we calculated the SNR versus magnitude with different filters under 30~Hz sampling, 27e$^{-}$ read-out noise, 1 arcsecond FWHM, $d=5$~counts/sec and no radio frequency interference. \fig{fig:snr} shows the relation between magnitude  and  SNR  for all filters.  Except for the y filter, stars with $m_{\rm V} < 11$ mag can achieve SNR $\ge 80$ with 30~Hz sampling. 

\section{Diffraction Profile of Stellar Occultation}
\label{sec:diff}
The differential size distribution $dN/dD \propto D^{-q}$ implies  that large objects are small in number. Therefore, the odds of serendipitous occultation 
by a large object is very small. Instead, we stand a better chance looking for occultations by objects in the small end of the size spectrum. There are two 
reasons for this: kilometer to sub-kilometer TNOs are far more numerous than big TNOs as the size distribution suggests, and the effective diameter of the occultation 
shadow is  enlarged due to the diffraction effect, which effectively increases the event rates. 

In diffraction theory, there are two regimes delimited  by the Fresnel number  $\tilde F=D^{2}/(r \lambda)$, where $D$ is the size of the TNO, $r$ is the distance 
from the object to the observer, and $\lambda$ is the observing wavelength: 
the {\em Fresnel} diffraction (near field) where   $\tilde  F  \gtrsim 1$ and {\em Fraunhofer} diffraction (far field) where $\tilde  F  \ll 1$. 
As we increase the size of the TNO or decrease the distance to the object, we pass from the Fraunhofer regime to the Fresnel regime. 
The characteristic size, known as  Fresnel scale $F$ ($\sim 2$~km at 43~AU, 600~$nm$),  is defined by $\sqrt{r \lambda}$ and can be used to identify the two regimes. 
The regime of  $ \tilde F \gg 1 $ is called the {\em geometric} regime, however for the scope of this work 
the probability of having such an occultation is insignificant.

\begin{figure}[]
\begin{center}
\includegraphics[width=90mm]{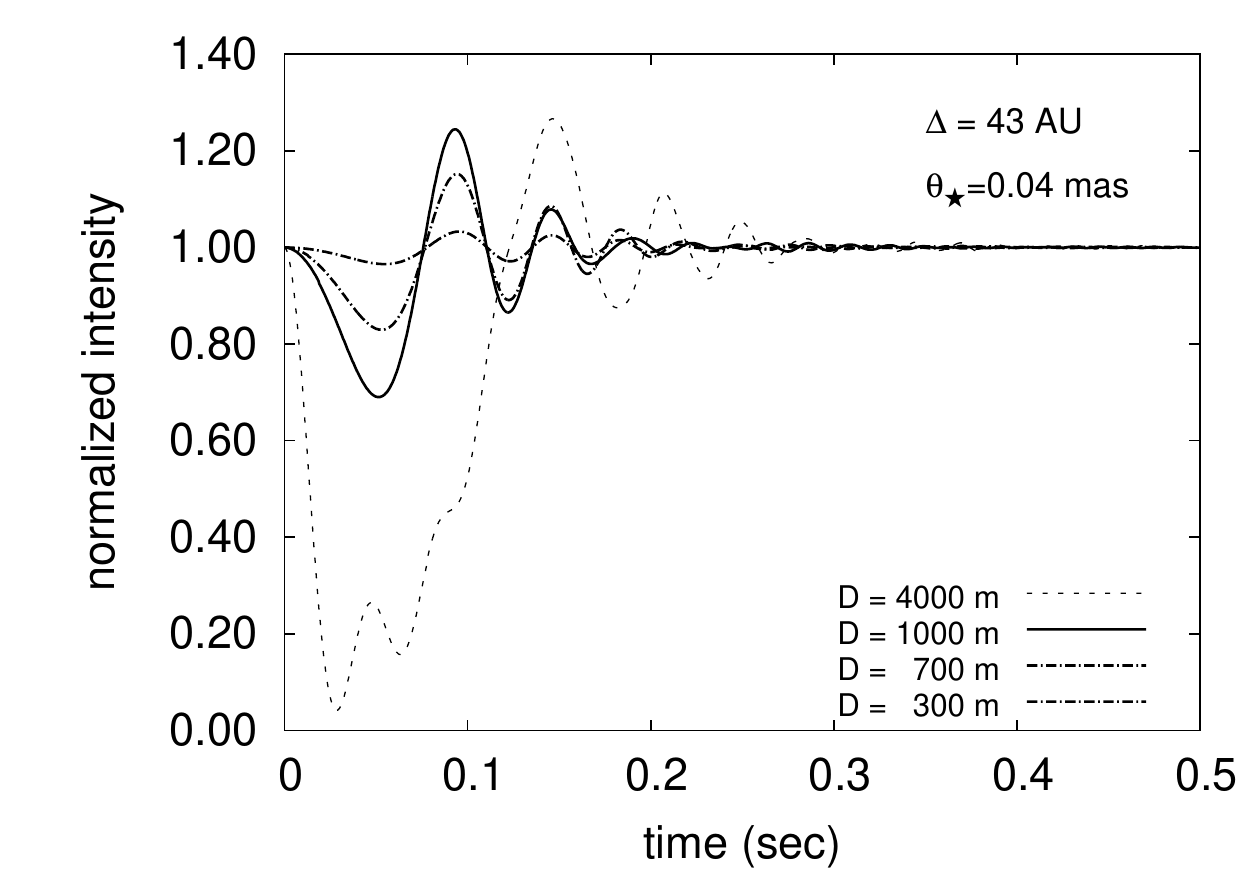}
\caption{Diffraction profiles for various TNO sizes at 43~AU and zero impact parameter at opposition. As the size of TNOs becomes smaller than the Fresnel scale, the effective diameter remains almost the same. In other words, for objects smaller than the  Fresnel scale, diffraction effect will make the effective diameter larger than the geometric diameter. The Fresnel scale at 43~AU is about 2~km, therefore in this case, 300~m, 700~m and 1000~m objects all have the same effective diameter.}
\label{fig:profile_size}
\end{center}
\end{figure}

The diffraction profile of stellar occultation for a point source can be calculated by the Lommel function and the profile for stars with finite size can be obtained by convolving the point source profile with the stellar disk \citep{Roques87}.  The  profile is a function of 
TNO distance, size and shape, stellar angular size and impact parameter  
 (defined as the distance between the line of sight of the star and the object). 
\citet{nihei2007} showed that the width of a diffraction profile $H$ (the diameter of the first Airy ring) with zero impact parameter 
is defined as:
\begin{equation}
\label{eqn:ed}
H(\lambda, r, D, \ts)=2\left[ ( \sqrt{3/2} F )^{\frac{3}{2}}+(D/2)^{\frac{3}{2}}\right]^{\frac{2}{3}}+ r \, \theta_{*} \, ,
\end{equation}
where  $\theta_{\star}$ is the angular size of occulted star.
\fig{fig:profile_size} shows diffraction profiles for various TNO sizes at 43~AU with zero impact parameter at opposition. As the size of TNOs becomes smaller than the Fresnel scale, the effective 
diameter remains almost the same. In other words, for objects smaller than the characteristic length defined by Fresnel scale, the diffraction effect will make 
the effective diameter independent of the size of the object. 

\begin{figure}
\begin{center}
\includegraphics[width=90mm]{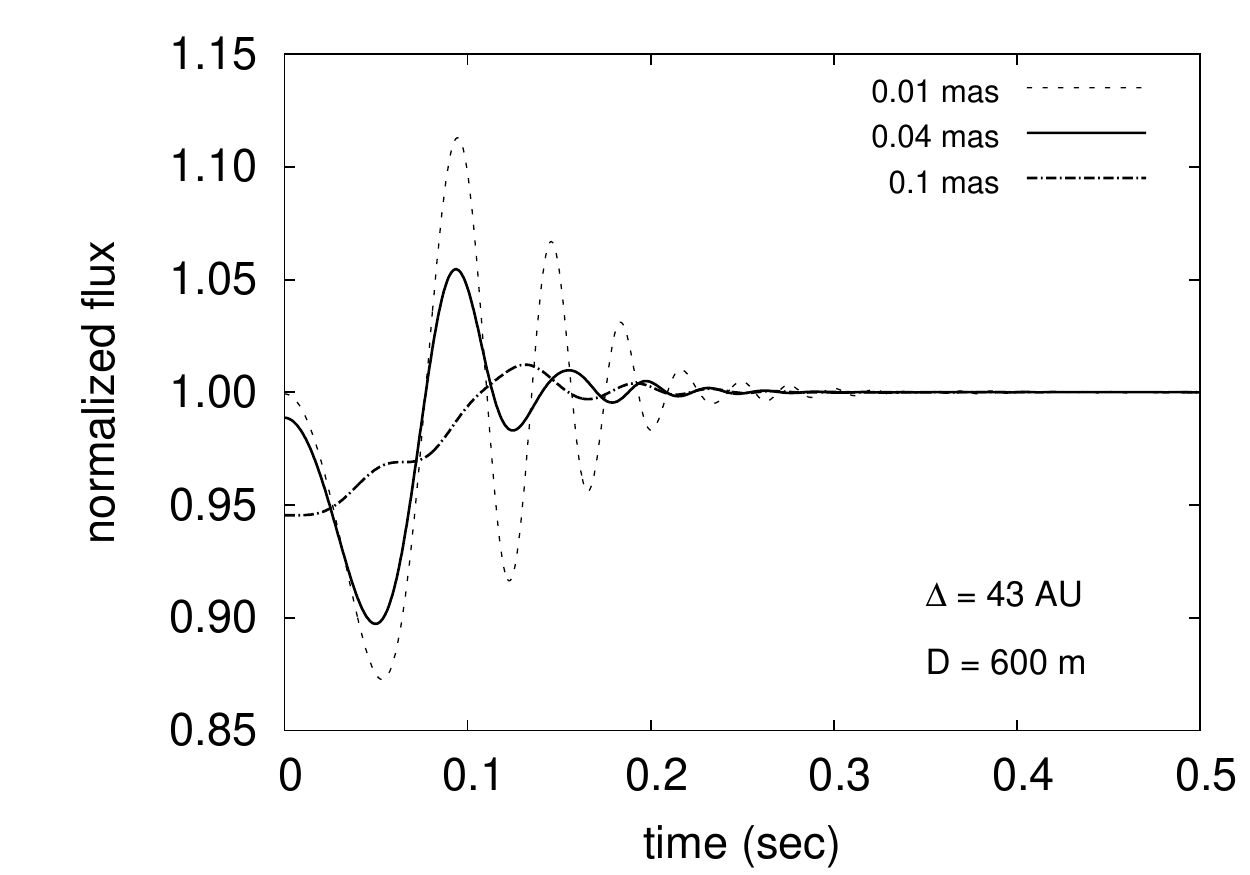}
\caption{Diffraction profiles of a 600~m size TNO at 43~AU and zero  impact parameter at opposition for various stellar angular sizes. 
Due to the fringe superposition, a star with large stellar angular size would smear out the diffraction profile and reduce the occultation depth. As it can be seen from the differences between 0.1 and 0.01~mas stars, the diffraction profile is very distinct for 0.01~mas star and becomes shallower when the stellar angular size gets larger.}
\label{fig:stellar}
\end{center}
\end{figure}

For non-point sources, due to the fringe superposition, stars with large stellar angular size would smear out the diffraction profile and reduce the depth of occultation as shown in  \fig{fig:stellar},  the diffraction profiles of 600~m size TNO at 43~AU for various stellar angular sizes. 

\begin{figure}[]
\begin{center}
\includegraphics[width=90mm]{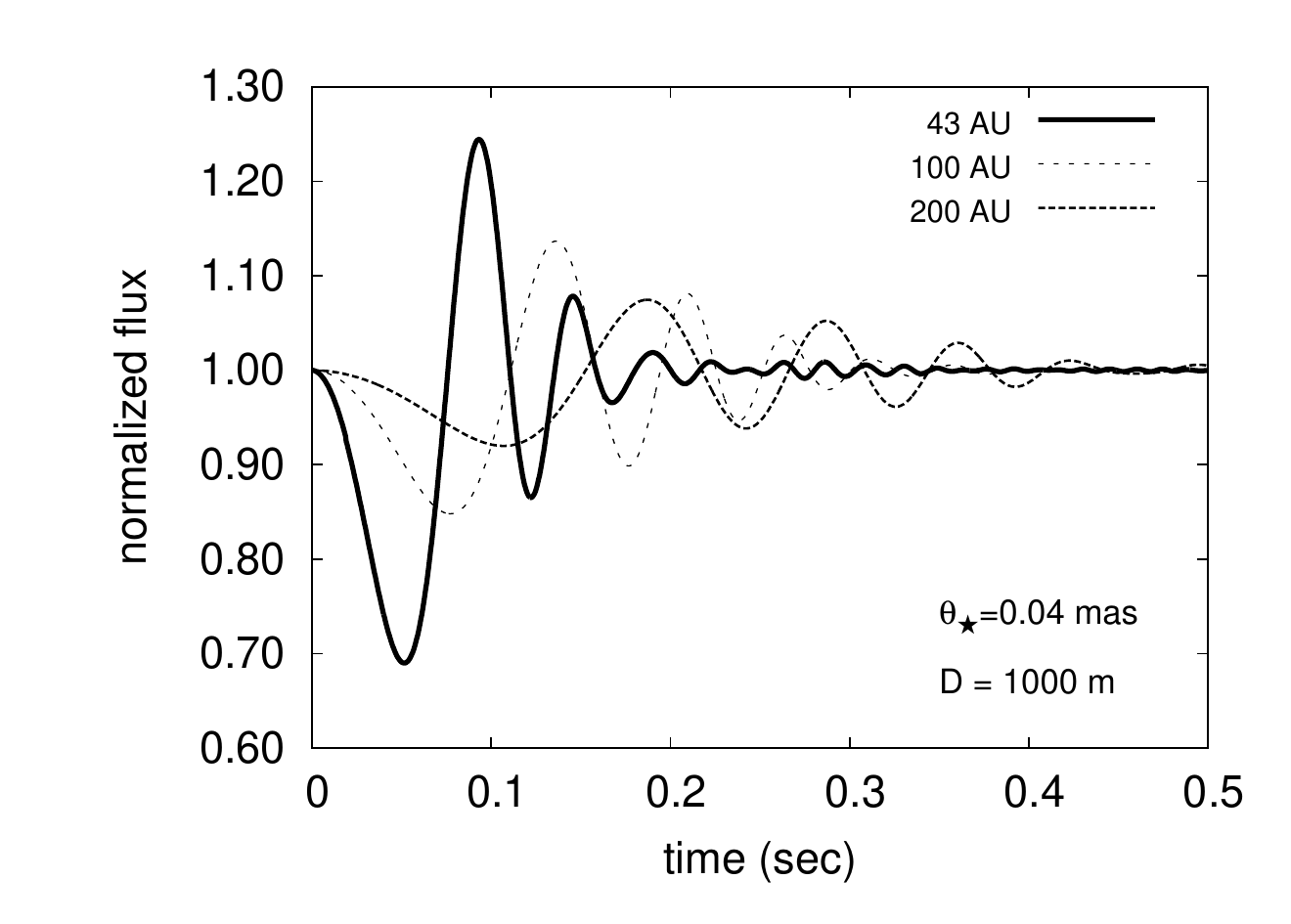}
\caption{Diffraction profile for 1~km size TNO at various distance with zero impact parameter at opposition. The effective diameter becomes larger and the depth of the flux reduction gets shallower as the distance to  the TNO increases.}
\label{fig:transit}
\end{center}
\end{figure}

For more distant objects, the Fresnel scale gets larger, and thus the same size objects 
are in the Fraunhofer regime. Furthermore, the projected 
size of the background star gets larger, thus a distant occultation event would have longer  
event duration but shallower occultation depth. In \fig{fig:transit}, as the distance 
to the objects increases, the effective diameter becomes larger, and the depth of the flux 
reduction gets shallower. 

The diffraction profile of \emph{spherical} objects have very distinct Poisson peak which can not be observed in irregular shape objects. However, as the objects become smaller than the Fresnel scale, the central Poisson peak exists for all shapes of objects \citep{Roques87}.

\begin{figure}[]
\begin{center}
\includegraphics[width=90mm]{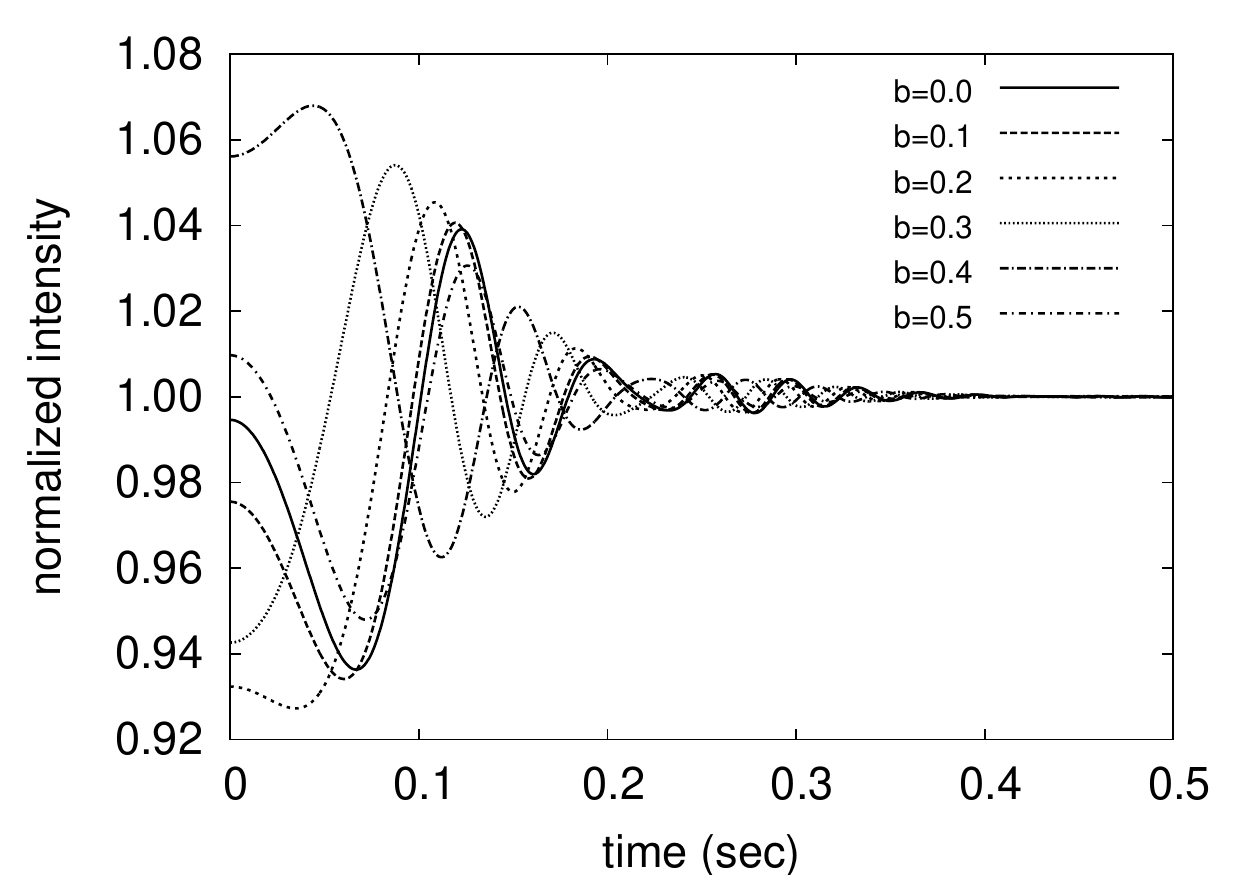}
\caption{Diffraction profile for 500~m size TNO at 43~AU with various impact parameters at opposition. The central Poisson peak drops as  the impact parameter increases until  it reaches the first Airy ring where the intensity goes up again.}
\label{fig:imp}
\end{center}
\end {figure}
Finally  Figure \ref{fig:imp} shows the diffraction pattern for various impact parameters. When the objects are  smaller than the Fresnel scale, the Poisson peak is prominent for $b=0$. The peak falls gradually, as $b$ increases, to a trough and then the intensity reverses (brightening event) as it reaches the first Airy ring.

One can notice that in the Fraunhofer regime,  
for example $D=500~\rm m$ at 43~AU, the shape and size do not change the width of the diffraction profile, only the depth of the occultation is affected by the object size and impact parameters. We based our detection method on this property as we explain in the 
following section.

\section{Detection Method}
\label{sec:det}
When the shadow of a stellar occultation sweeps across a telescope, 
high sampling rate systems like PS1 video mode will translate the shadow with diffraction profile 
into several time-sampled lightcurve points.  In order to look for  signatures of occultation, a detection method 
should be able to probe these lightcurves for  diffraction patterns.
There are far more numerous TNOs smaller than the Fresnel scale, 
therefore we target our search  at objects with $D \leq F$. 
In this regime, $H$ does not  change as a function of the object size, shape or stellar spectral type. 
On the other hand, the depth  of the occultation  depends on the   
object size and stellar angular size. Also,
 the shape of the diffraction profile  depends on the impact parameter. 

Based on these conditions, we have designed our detection method  to search for occultations by 
scaling  pre-defined TNO diffraction profiles (i.e. templates) to match patterns in the lightcurves. 
For a given star, with known stellar angular size $\theta_{\star}$ and SNR $S$,
let $g(\tau; b, \{  \theta_{\star}, S\})$ be a normalized template with mean zero
as a function of time $\tau$ and impact parameter $b$.
We then try to find the best matched template by maximizing the following measure,
\begin{equation}
	{\cal M}(\tau)  =  \sum_{\tau=t-w}^{\tau=t+w}  \left[\alpha \, g(\tau; b, \{  \theta_{\star}, S\}) - p(\tau)\right]^{2}  \,\, ,
\end{equation}
with respect to the scaling factor $\alpha$, where  $p$ is the local nominal baseline  subtracted lightcurve within a window 
$ \left[-w : +w\right]$. The size of the window $n$ can be found analytically as $sH/v_{\rm rel}$, 
where $s$ is the sampling rate, $v_{\rm rel}=v_{\rm e} \,(\cos \phi-\sqrt{r^{-1}_{_{\rm AU}}})$
is the relative velocity with $v_{e}$  is the orbital speed of the Earth, $\phi$ is the opposition angle and $r_{_{\rm AU}}$ 
is the distance of a TNO to the observer in units of AU. 
Finding the scaling parameter $\alpha$ can be done analytically as:
\begin{equation}
	\label{eq:alpha}
	\alpha(b, \{\ts,S \})  =  \frac{\sum {p}(\tau)  \,g(\tau; b, \{  \theta_{\star}, S \} ) } {\sum g(\tau; b, \{  \theta_{\star}, S\})^{2}} \,.
\end{equation}

\noindent To  consider the effect of  $b$, we simply maximize $\alpha$ 
over $b$,

\begin{equation}
	\label{eq:alpha_max}
	\tilde \alpha( \{\ts, S\})  =  \max_{b} \left[  \alpha(b, \{\ts, S\})  \right]  \, .
\end{equation}
In practice,   for a given $\ts$ and $S$, we built a set of templates with impact parameter  values,
$b=\{0.00H, 0.01H, 0.02H, \dots, 0.50H \}$,
and scan all templates across the lightcurve to find the template that maximizes $\alpha(b, \{\ts, S\})$.
In order to speed up the computation, we actually perform a two-pass procedure. 
The first pass involves scanning corresponding templates from
$b=0.0H$ to $0.5H$ with $\Delta b=0.1H$ to find the region that brackets the 
maximum $\alpha$. Then within this region, we apply a second pass 
that uses templates with finer impact parameter step, $\Delta b=0.01H$, and again 
find the maximum $\alpha$ value within this region. The template that 
gives the largest $\alpha$ value is the best fit between the  scaled
template and a pattern in the lightcurve. At the end of this process, we  have the same number of optimized 
$\alpha$'s as the number of points in all lightcurves\footnote{There are actually fewer  $\tilde \alpha$'s  than number of points 
because at the beginning and end of the lightcurves there are not enough points to define the local nominal baseline.}.
In order to decide if any of these $\alpha$ values represent real events 
or false positives, we determined a threshold value $\alpha_{\rm{t}}(\ts,S)$,  calculated from  the \emph{null} hypothesis distribution. 
For this purpose, 
we simulated lightcurves with no events, then scanned templates through these lightcurves, optimized for $\alpha$'s,  built the  
distribution, set the  p-value based on the  desired false positives  and finally determined the $\alpha_{\rm{t}}(\ts,S)$ accordingly.

This distribution can also be estimated semi-analytically, thereby reducing 
the computational cost. Finding an optimized parameter using  a least square 
metric  while minimizing over another parameter 
is equivalent to a regression with one nuisance parameter \citep{rice2007}.  The cumulative distribution
of such a system is known to  follow the \emph{Gumbel distribution} or type I \emph{extreme value distribution} 
\citep{gumbel1943},
\begin{equation}
	\label{eq:evd1}
	G(X \leq x)=e^{-e^{-\nu(x-\mu)}}\,\,,
\end{equation}
where $\nu$ and $\mu$ are fitted parameters (scale and location parameters). 

\begin{figure*}[]
\begin{center}
\begin{tabular}{cc}
\hspace{-.8cm}
\resizebox{90mm}{!}{\includegraphics{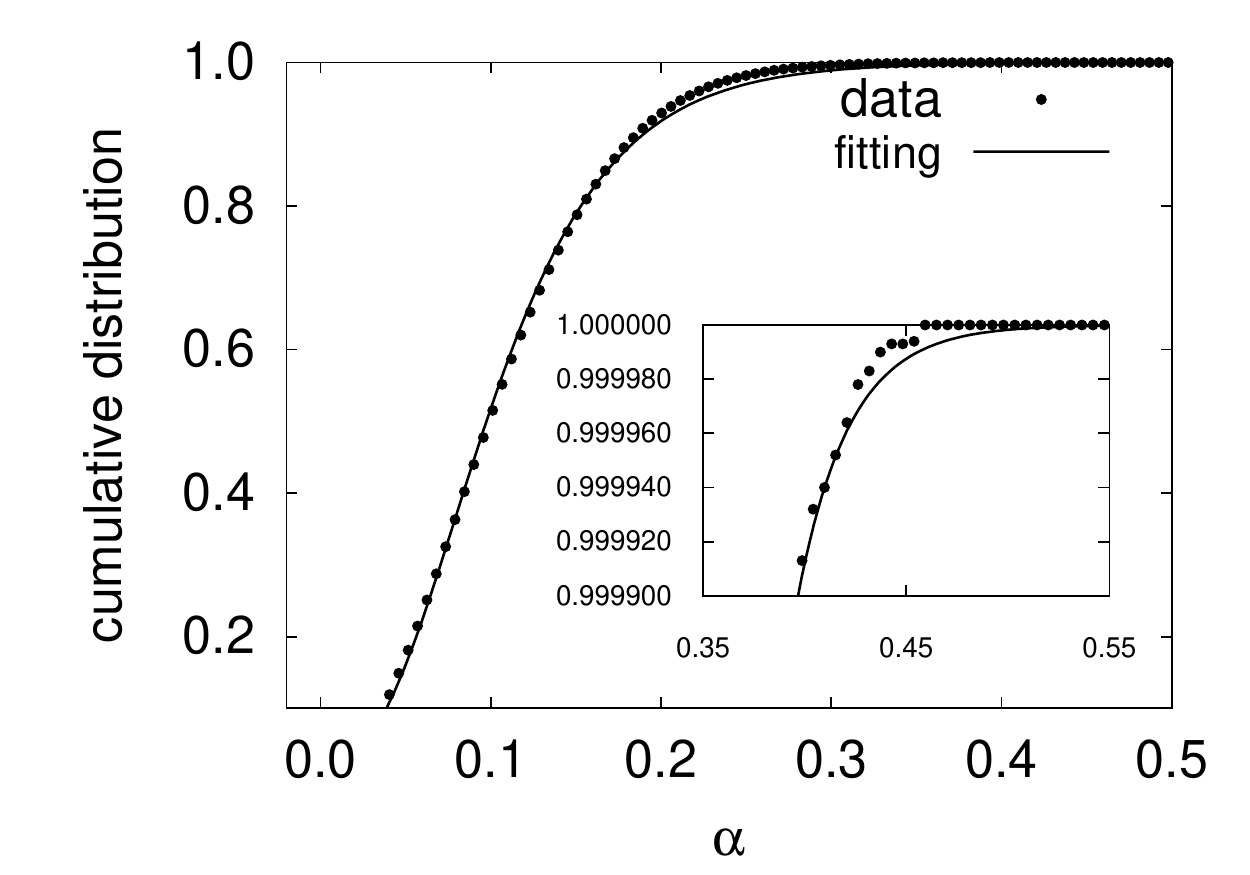}} &
\hspace{-1.3cm}
\resizebox{90mm}{!}{\includegraphics{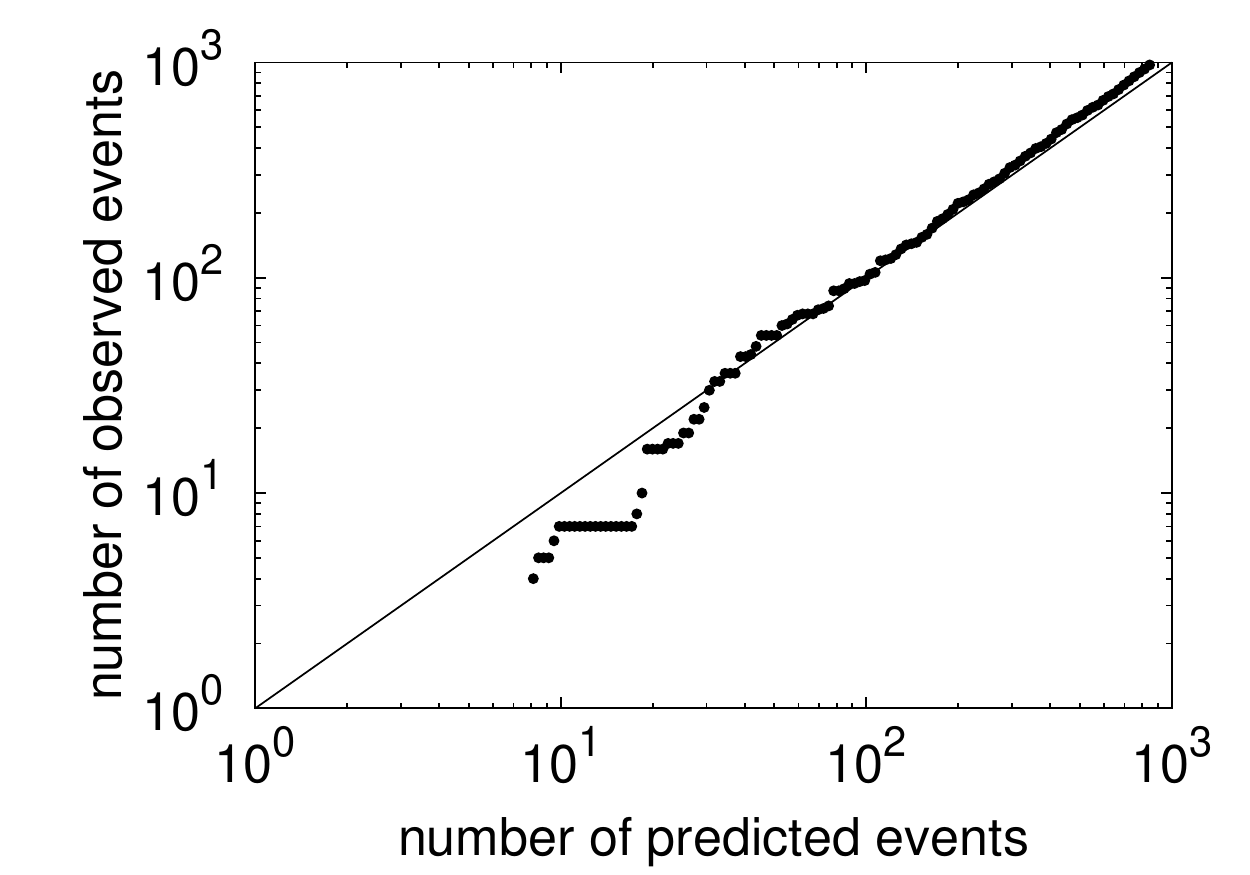}} \\
\end{tabular}
\caption[]{Left panel shows the null hypothesis distribution of $\alpha$ values (shown for SNR=100 $\ts=0.01$~mas with dots) and fitted Gumbel distribution (solid line). From a separate fitting to the tail part (inset), the difference between the null hypothesis distribution from simulation and the fitted Gumbel distribution is very small. The right panel shows the number of false positives predicted by Gumbel distribution versus the number of false positives observed in the simulation. The diagonal line is to show the difference between predicted and observed values. As can be seen that at high threshold (bottom-left corner), the number of predicted false positives is more than the number of observed false positives.}
\label{fig:nulldist1}
\end{center}
\end{figure*}

The left panel of \fig{fig:nulldist1} compares the null hypothesis  cumulative distribution of $\tilde \alpha$'s,  for  SNR=100 and $\theta_\star=0.01$~mas, 
built from simulations (solid points) with the fitted Gumbel distribution  (solid line). 
As can be seen from the figure, the fitting for cumulative distribution $< 0.8$ ($\tilde \alpha < 0.2$) was good. But the fitted part of the tail  ($\tilde \alpha > 0.2$) of the distribution, which is the part  we care about the most, did not fit well. 
Therefore, we separately fitted the tail part of the cumulative distribution, and the results are shown in the inset of left panel of the same figure. 
Note that the difference in the tail between the null hypothesis distribution and the fitted Gumbel distribution is now very small. The right panel of \fig{fig:nulldist1} shows the number of false positives predicted by Gumbel distribution versus the number of false positives observed in the simulation. As can be seen, at high threshold (bottom-left corner) the number of predicted false positives is more than the number of observed false positives. This is due to the fact that the $\alpha_{\rm{t}}(\ts,S)$ set by fitted Gumbel distribution is actually higher than that of the true null hypothesis distribution. 
As a result, even though the threshold will be set a bit higher,
$\alpha_{\rm{t}}(\ts,S)$ can be still  calculated based on the Gumbel distribution  
to ensure one false positive in the lifetime of the project. In reality, we may get 
fewer than one false positive, as we demonstrated above.

The next step is to search in the lightcurves for any
$\tilde\alpha$ values that are larger than  the $\alpha_{\rm{t}}(\ts,S)$. Note that the 
$\alpha_{\rm{t}}(\ts,S)$ is $\ts$ and $S$ dependent, thus the process of simulating
the null hypothesis distribution and fitting for $\nu$ and $\mu$  has to be repeated for each individual star.  
In order to alleviate this problem, we have constructed a grid of fitted $\alpha_{\rm{t}}(\ts, S)$'s 
values for various  SNRs and angular sizes a priory and tabulated these values into a database.
 For a given star, the  threshold is approximated with  $\alpha_{\rm{t}}(\ts', S')$ where $\ts'$ and $S'$
 are the closest pre-calculated values to $\ts$ and $S$.

Note that in the presence of auto-correlations (trends) in the lightcurves the 
Gumbel distribution may not best describe the data. We have not yet investigated how well this approximation performs when trend exists in the lightcurves. 
Therefore, in the event detection pipeline, we  apply a high-pass filter and de-trend the lightcurves to avoid this difficulty (see $\S$\ref{sec:engdata}).
   
\begin{figure*}[]
\begin{center}
\begin{tabular}{cc}
\hspace{-.8cm}
\resizebox{90mm}{!}{\includegraphics{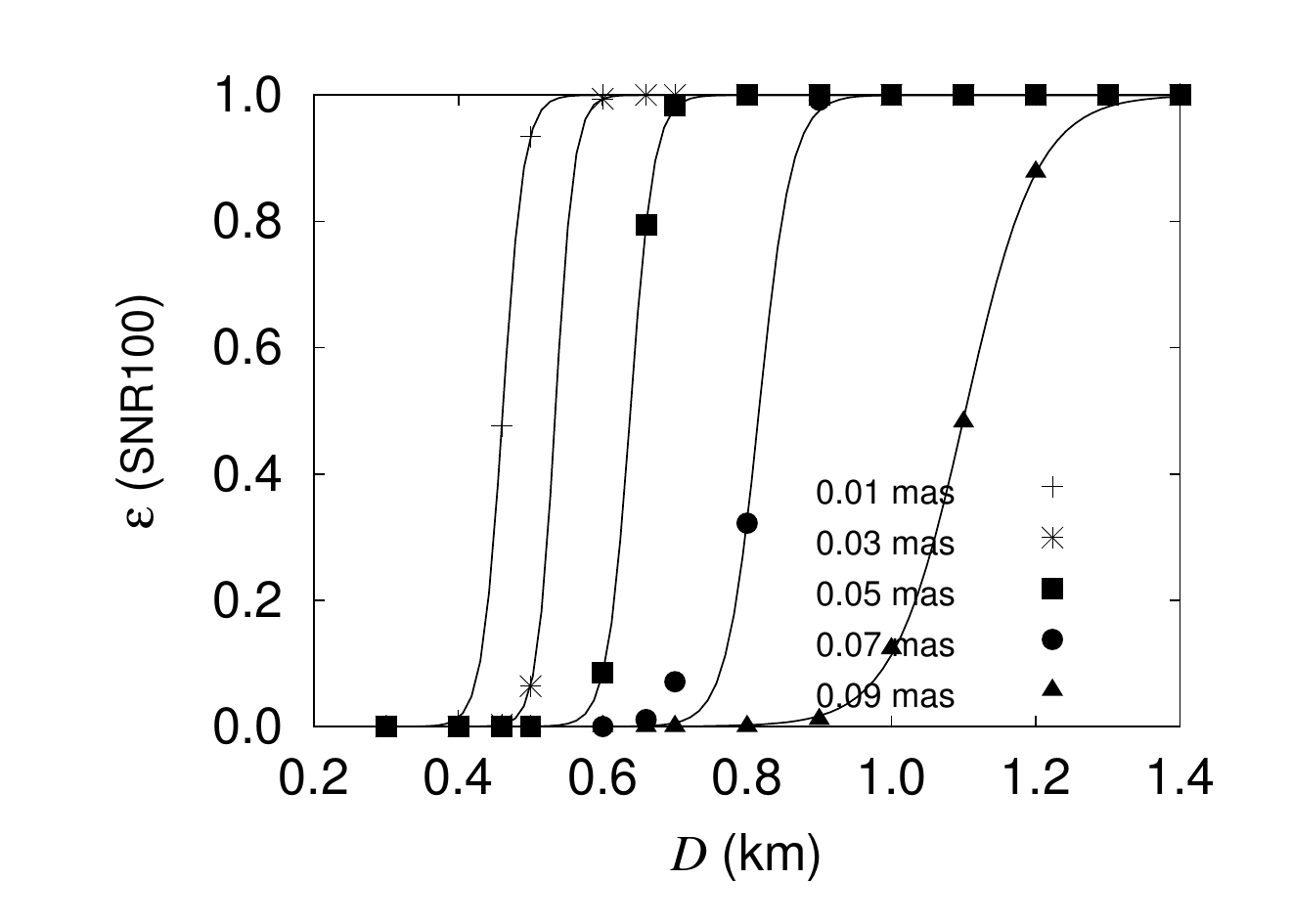}} &
\hspace{-1.2cm}
\resizebox{90mm}{!}{\includegraphics{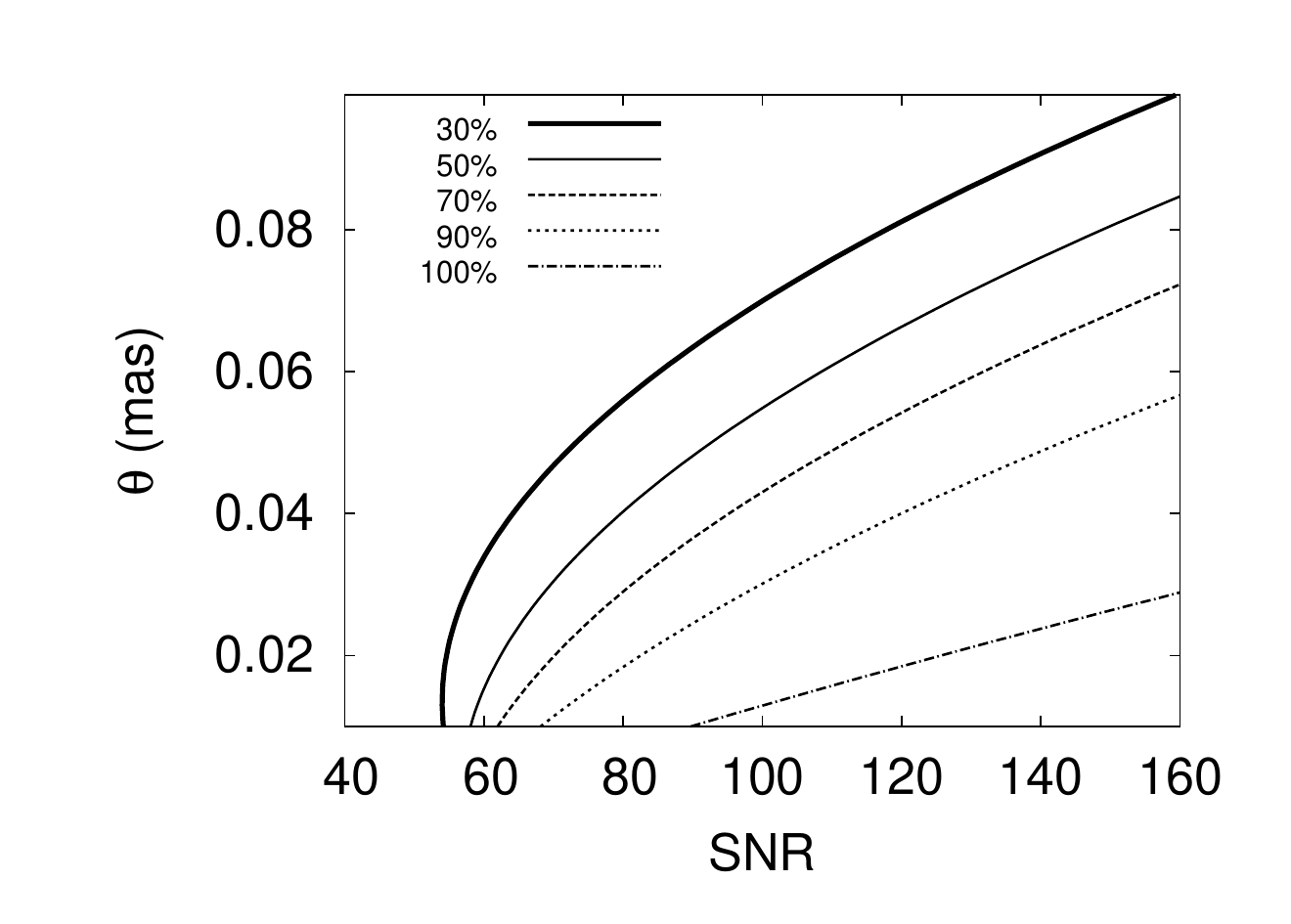}} \\
\end{tabular}
\caption[]{The left panel is the efficiency as a function of TNO size for SNR=100 and various stellar angular sizes. The right panel is the efficiency contour as function of
SNR and stellar angular size for 700 m objects. Stars with higher SNR and smaller stellar angular size have higher detection efficiency and are best candidates for searching 
stellar occultation in the lightcurves.}
\label{fig:efficiency}
\end{center}
\end{figure*}

To estimate the detection efficiency as a function  of TNO-size for PS1 guide stars with 
various SNRs and angular sizes,  we implanted objects with various sizes into 
simulated lightcurves and calculated the number of $\tilde \alpha$ values for each object size that are larger than the  threshold value $\alpha_{\rm{t}}(\ts, S)$. 
\fig{fig:efficiency} left panel shows the detection efficiency for different stellar angular sizes as a function of 
object size. As the plot suggests, PS1 can detect objects as small as  $\sim 400$~m. 
\fig{fig:efficiency} right panel shows the detection efficiency as a function of SNR and $\ts$ for 700 m obejcts; the detection 
efficiency is higher for higher  SNRs and smaller stellar angular sizes.
Therefore the guide stars selection (see $\S$\ref{sec:catalog})  should be  based on this fact: 
stars with higher SNR and smaller stellar angular size are best candidates for searching stellar occultation in the lightcurves. 

To evaluate whether our detection process is 
computational feasible, we have estimated the total float point operations (FLOP) needed 
to process all data for  one night to be  $\sim 10^{11}$ (this is a conservative estimate). To save computational time, 
we firstly tabulate the values of   $\sum g(\tau; b, \{  \theta_{\star}, S\})^{2}$ in \eq{eq:alpha}. 
 The two-pass procedure to search for matched template needs about 25 
FLOP. Each template has 13 data points, therefore for each data point in the lightcurve, the computational cost is about 325 FLOPS
and for each night, there will be  a total of $\sim 5\times 10^{7}$ data points assuming 30Hz sampling and 60 guide stars. 
With a 3G~Hz CPU, the whole detection process can be done in few minutes. 

\cite{bickerton2008} has demonstrated a similar detection method using convolution 
of templates with lightcurves. The advantage of convolution done in Fourier space is that 
it is computationally faster. Nevertheless,  our computational cost is not significant
as we have pointed out above, and therefore we choose to do the template matching in real space
in order to avoid problems of aliasing and for better control and understanding of the 
underlying distribution. 

\section{Estimate of The Stellar Angular Size}
\label{sec:angular}
As mentioned before, the stellar angular size is  important in choosing the 
guide stars.  For this reason we need to be able to estimate these
angular sizes  ahead of time in order to select guide stars that will yield the highest
number of events.  

\begin{figure}[]
\begin{center}
\begin{tabular}{cc}
\hspace{-.8cm}
\resizebox{90mm}{!}{\includegraphics{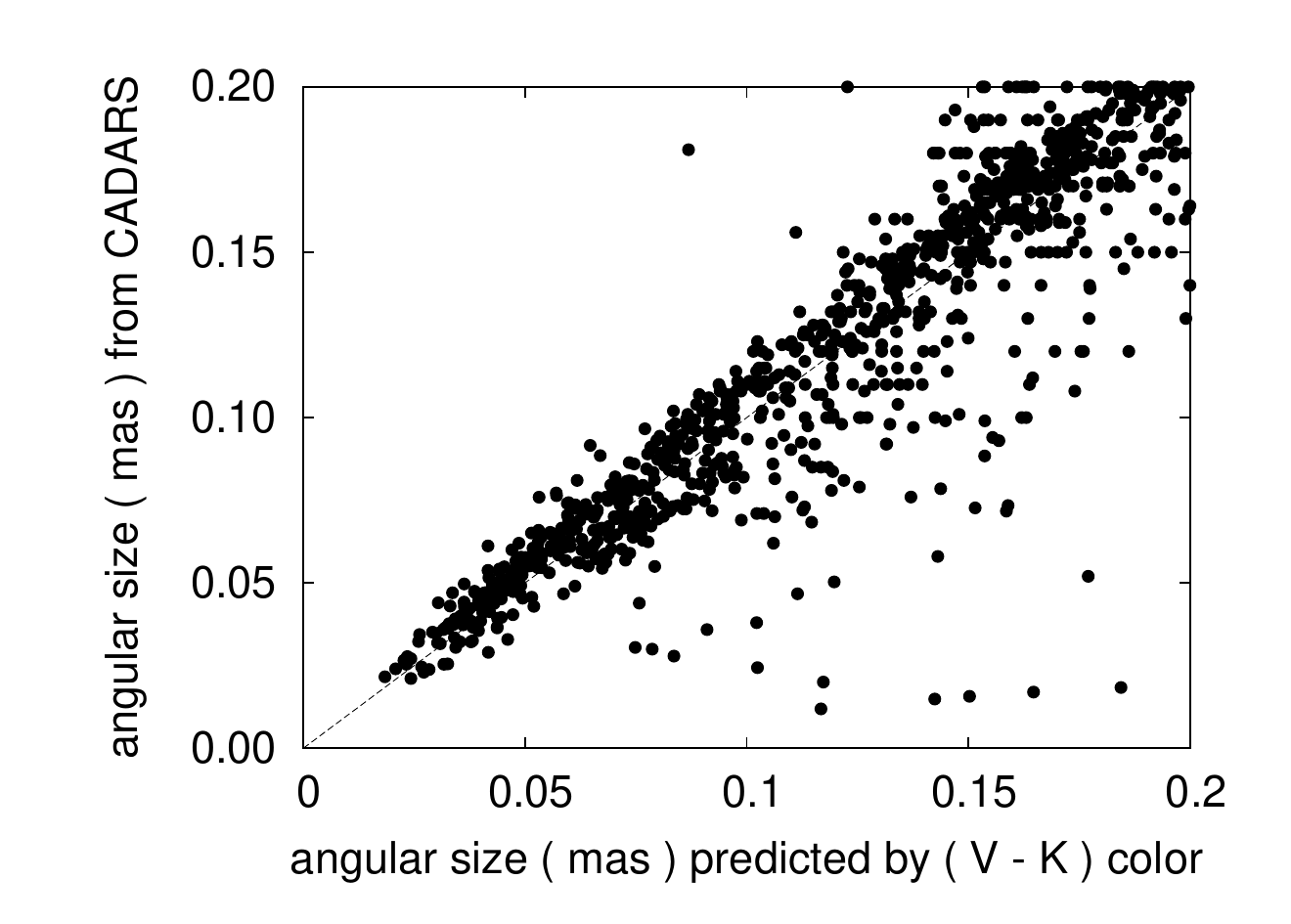}} &
\hspace{-1.2cm}
\resizebox{90mm}{!}{\includegraphics{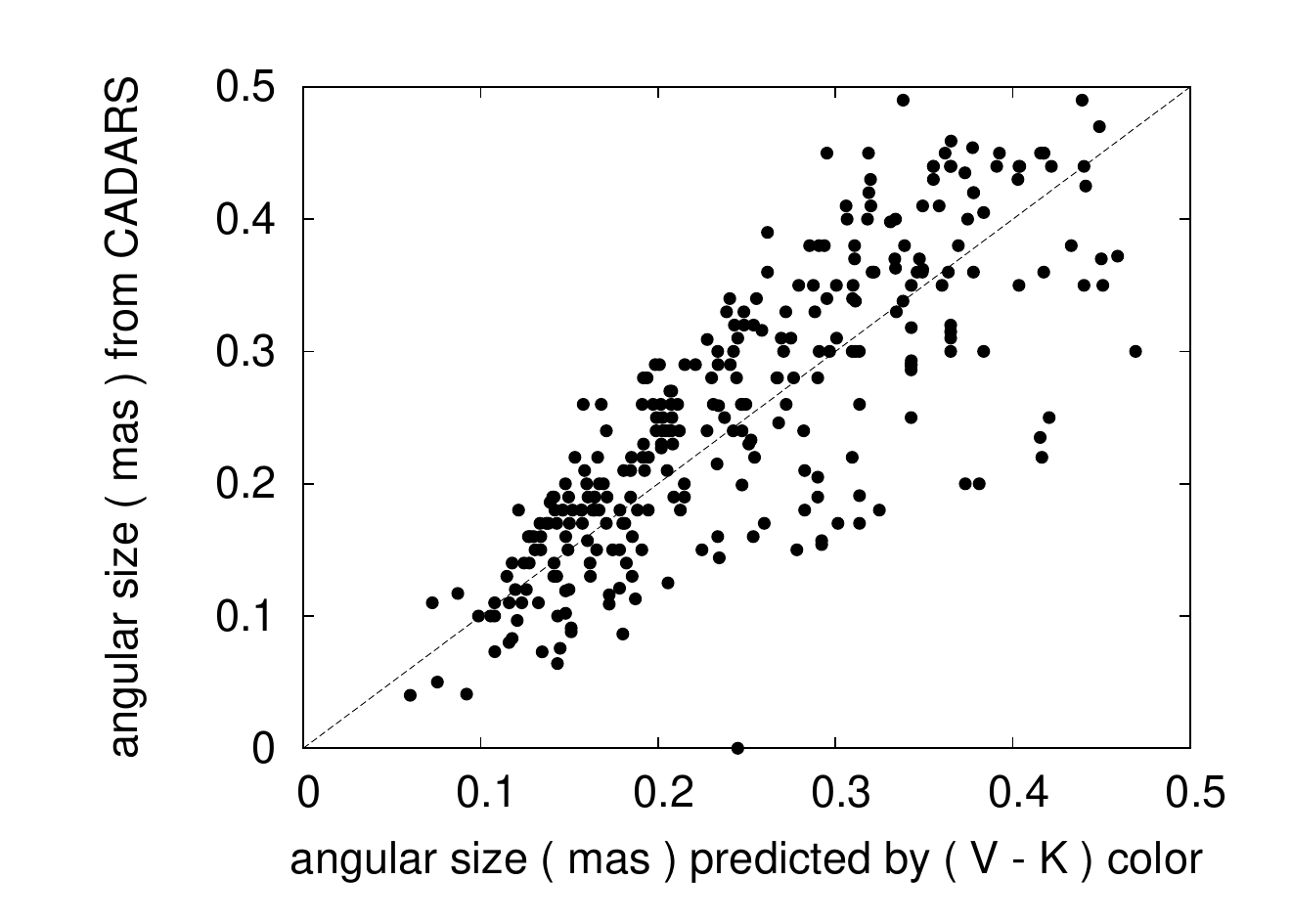}} \\
\end{tabular}
\caption[]{Angular sizes predicted from \eq{equ:ms} and \eq{equ:giant}.  The predicted values matched well with both  main sequence (left panel) and giant (right panel) stars from CADARS.}
\label{fig:size}
\end{center}
\end{figure}

Various methods have been used  to measure stellar angular sizes -for example,  lunar occultation, stellar interferometry, speckle interferometry, and other  
photometric and spectrophotometric methods (see \citet{scholz1997} and reference therein). 
However, these measurements from the literatures are limited to relatively few bright stars. 
An alternative way to select the most appropriate  guide stars from all stars visible to PS1 $3\pi$ sky, 
is to estimate the stellar angular sizes using their colors.
For example, \citet{vanbelle1999} fitted a few stars with known zero magnitude angular sizes ($\theta_{V=0}$ and $\theta_{B=0}$)  versus their $(V-K)$ and $(B-K)$ colors to set up the relation between the color indices and their angular sizes. Accordingly, 
we used the formulae in his work in combination with  the  
Tycho2\,\footnote{http://archive.eso.org/ASTROM/TYC-2/readme.htm} catalog \citep{tycho2} for $V$-band magnitudes,  
and the 2MASS\,\footnote{http://www.ipac.caltech.edu/2mass/} catalog  \citep{2mass}  for $K$-band magnitudes.
We first extracted all stars with $m_{\rm V}< 13$ mag and $m_{\rm K}< 16$ mag above the PS1 southern declination limit 
$\delta \ge -30^{\circ}$ and cross matched stars from each catalog by looking for pairs within one arcsecond radius
 and calculated their corresponding ($V-K$) color.
 We then refitted recent measurements of angular sizes from the Catalog of Apparent Diameter and Absolute Radii of Stars (CADARS) compiled by \citet{Pasinetti2001}  with \citet{vanbelle1999}'s relations. We  obtained the following fitted relation: 
\begin{equation}
\theta=10^{\left[0.453\pm0.003+0.246\pm0.005(V-K)-0.2V\right]} \, ,
\label{equ:ms}
\end{equation}
\noindent for main sequence stars and 
\begin{equation}
\theta=10^{\left[0.407\pm0.007+0.238\pm0.009(V-K)-0.2V\right]} \, ,
\label{equ:giant}
\end{equation}
 for giants.
 The fitting results are shown in \fig{fig:size} and as  can be seen, the predicted values matched well with the measured ones for both main sequence and giant stars. 

\begin{figure}[]
\begin{center}
\begin{tabular}{ccc}
\hspace{-.8cm}
\resizebox{64mm}{!}{\includegraphics{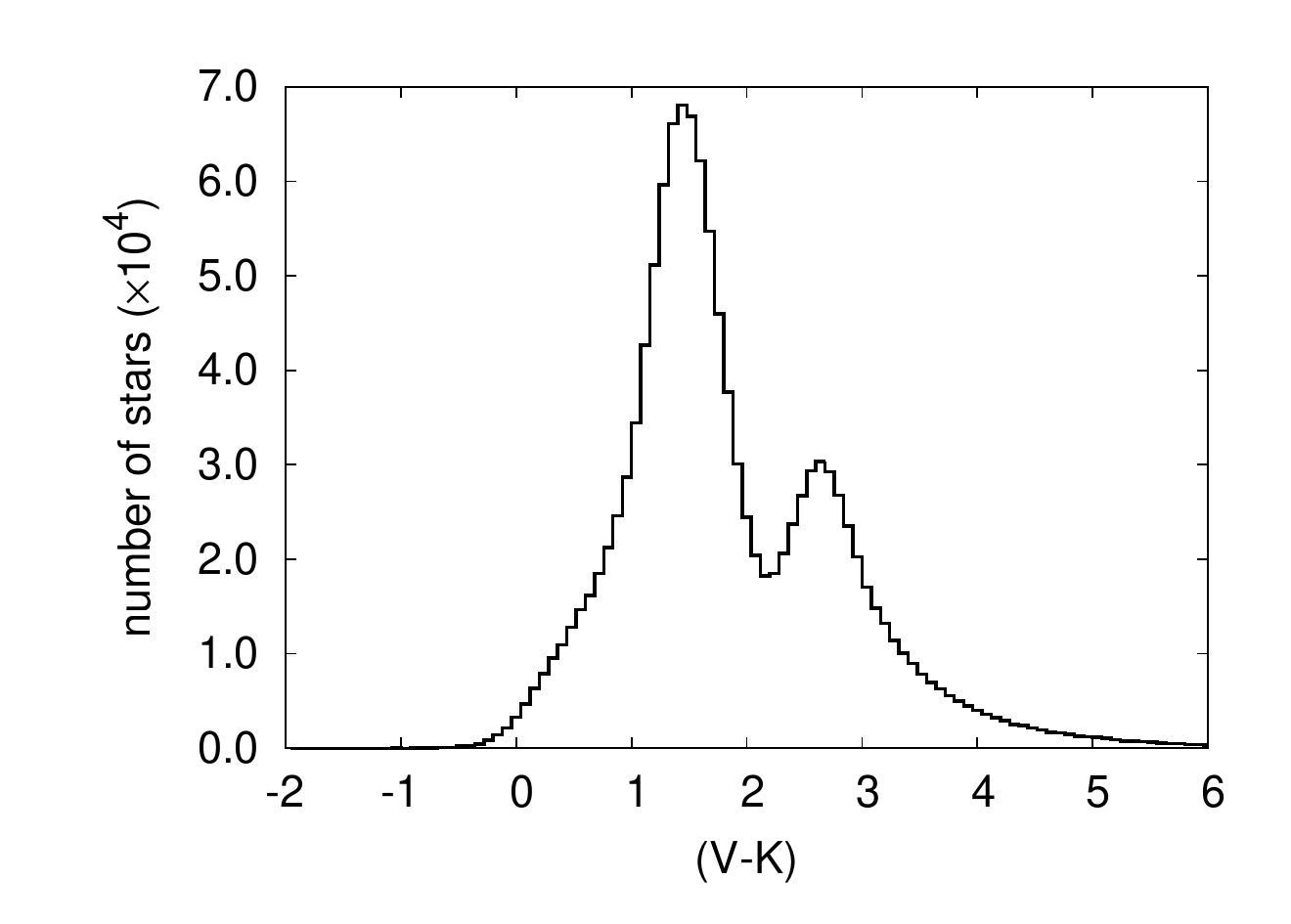}} &
\hspace{-1.4cm}
\resizebox{64mm}{!}{\includegraphics{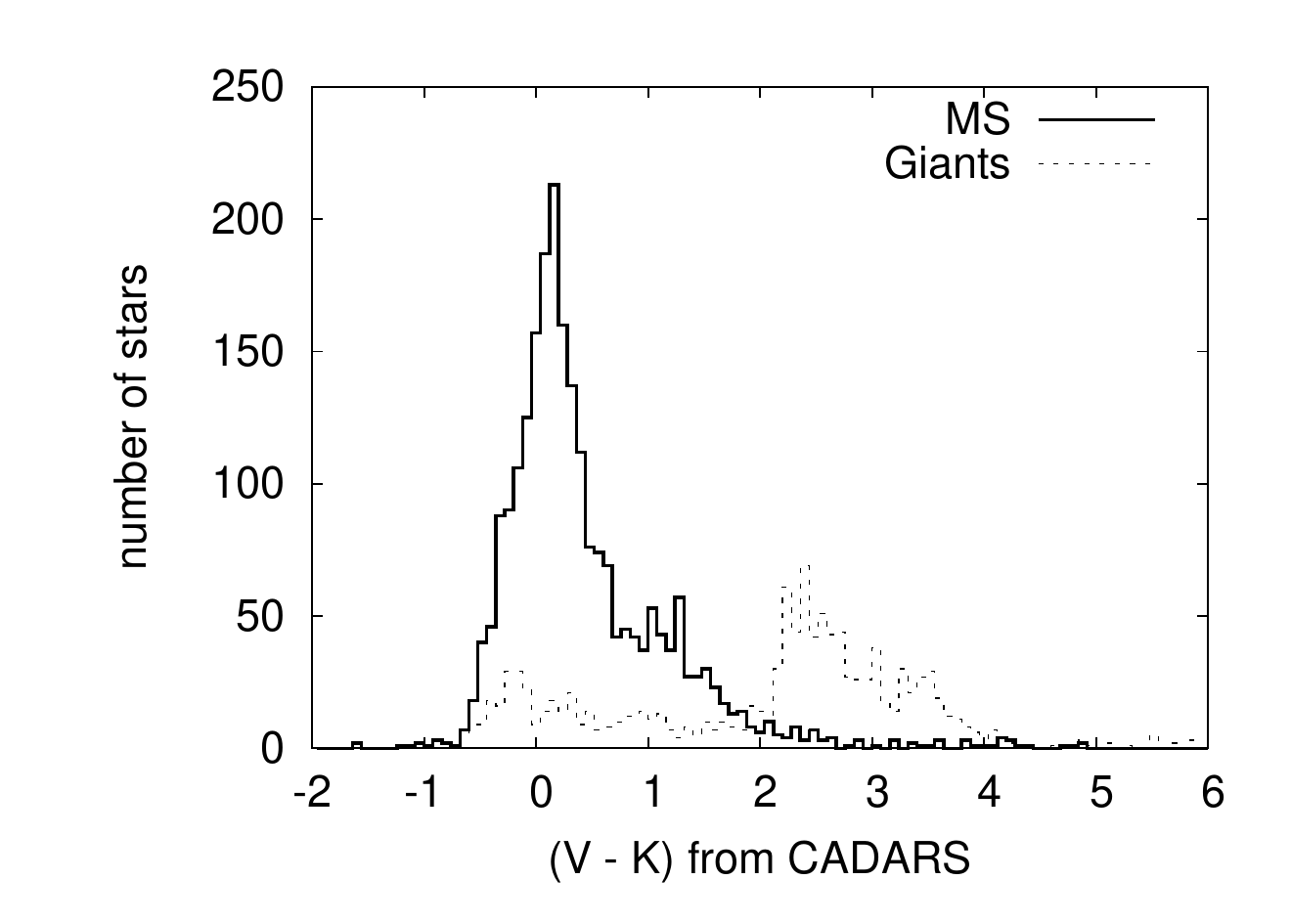}} &
\hspace{-1.4cm} 
\resizebox{64mm}{!}{\includegraphics{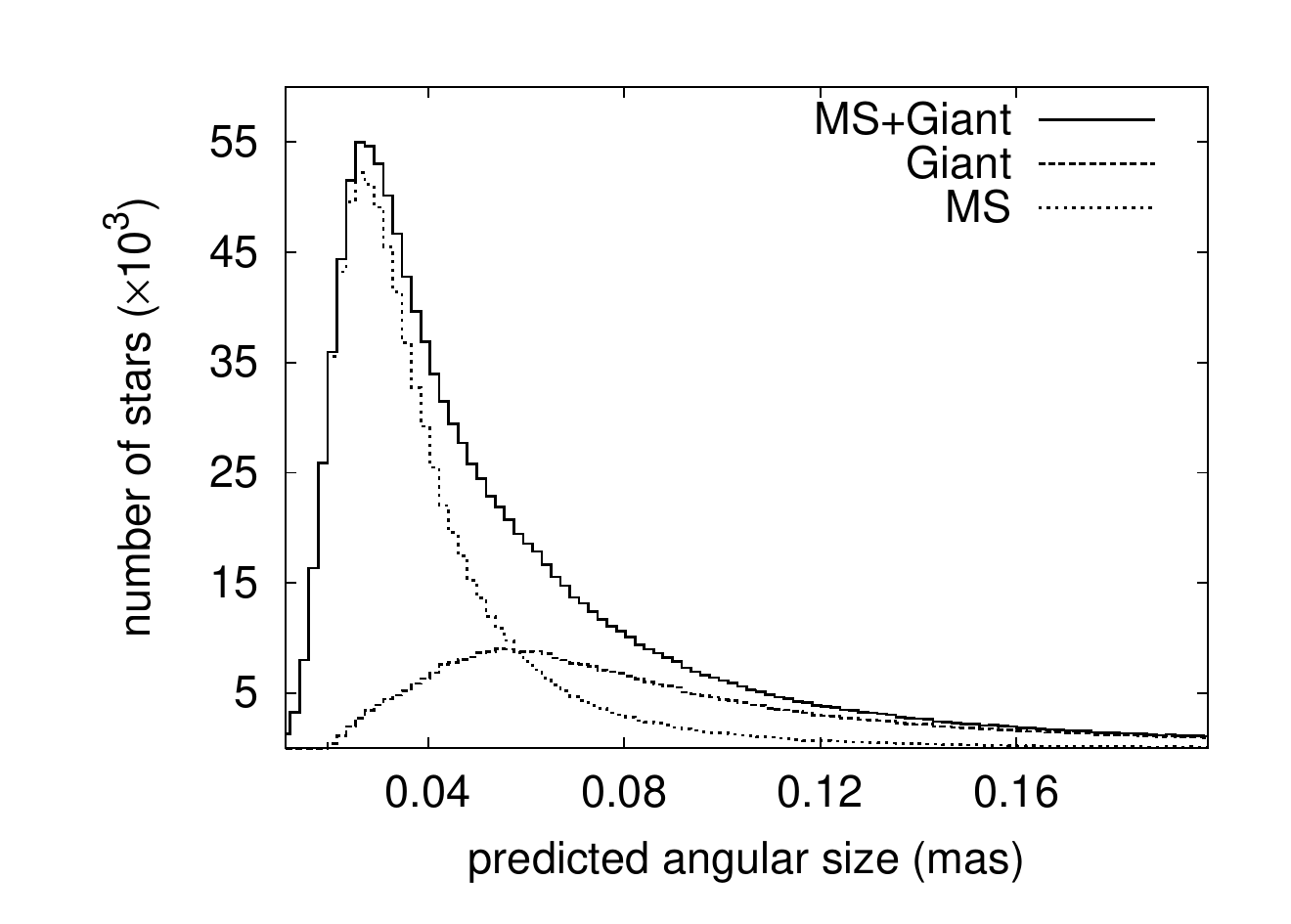}}  \\
\end{tabular}
\caption[]{Left panel: Distribution of $(V-K)$ of all matched stars between Tycho2 and 2MASS catalogs. As it can be seen, there are two peaks in the distribution, indicating contribution from main sequence (left peak) and giant (right peak). Middle panel:  The $(V-K)$ distribution of the measurements from CADARS. The number of main sequence stars dropped quickly for $(V-K)>2.0$ and so is the number of giants for $(V-K)<2.0$.  Right panel: The distribution of predicted angular size of the matched stars between Tycho2 and 2MASS. Main sequence stars peak around 0.02~mas, and giants peak around 0.05~mas. The overall distribution (solid-line) peaks around 0.02~mas.}
\label{fig:vmk}
\end{center}
\end{figure}

Due to the fact that stellar angular sizes depend not only on the spectral type but also on the luminosity class, the next question is how to reckon the  luminosity class  for every matched star when there is no such information. 
To do so we relied on the $(V-K)$ color as a way to (roughly) distinguish between main sequence and giant stars. 
On the left panel of Figure \ref{fig:vmk} we show that  the distribution of $(V-K)$ from all matched stars in PS1 3$\pi$ sky  has two peaks,  suggesting two groups of stars - most likely main sequence  and giants.  We used the luminosity class information provided in CADARS  to determine the boundaries 
between these two classes (shown in Figure \ref{fig:vmk} middle panel). Based on this, all stars with $(V-K)\leq1.85$ were considered to be the main sequence stars and all stars with $(V-K)>2.0$ to be giants. We abstained from classification for stars between these boundaries. 
Note there are still  a few giant stars with small $(V-K)$ (see middle panel of \fig{fig:vmk}). This is a selection bias in the catalog due to the fact that these are bright blue O or B giant stars with multiple measurements. In reality, O and B giant stars are rare and can be removed from the final catalog 
using known O and B star catalogs \citep{reed2003, maiz2004}. 

Using the $(V-K)$ cut,  \eq{equ:ms} and \eq{equ:giant}, we predicted angular size for CADARS stars and estimated the rms error to be 21.7\% for main sequence and 20.6\% for giants.  

Finally we estimated the angular sizes for all matched stars visible to PS1 3$\pi$ sky  and the resulting distributions are shown in  \fig{fig:vmk} (right panel). 
 Main sequence stars peak around 0.02~mas, and giants peak around 0.05~mas where the overall distribution (solid-line) peaks around 0.02~mas.

\section{Event Rate Based on Various Models}
\label{sec:rate}
In this section we estimate the number of expected events for the
three year project lifetime based on different models.  The comparison of 
the number of expected events with the actual observed events will 
allow for discrimination between the different models. 

\begin{figure}[]
\begin{center}
\includegraphics[width=90mm]{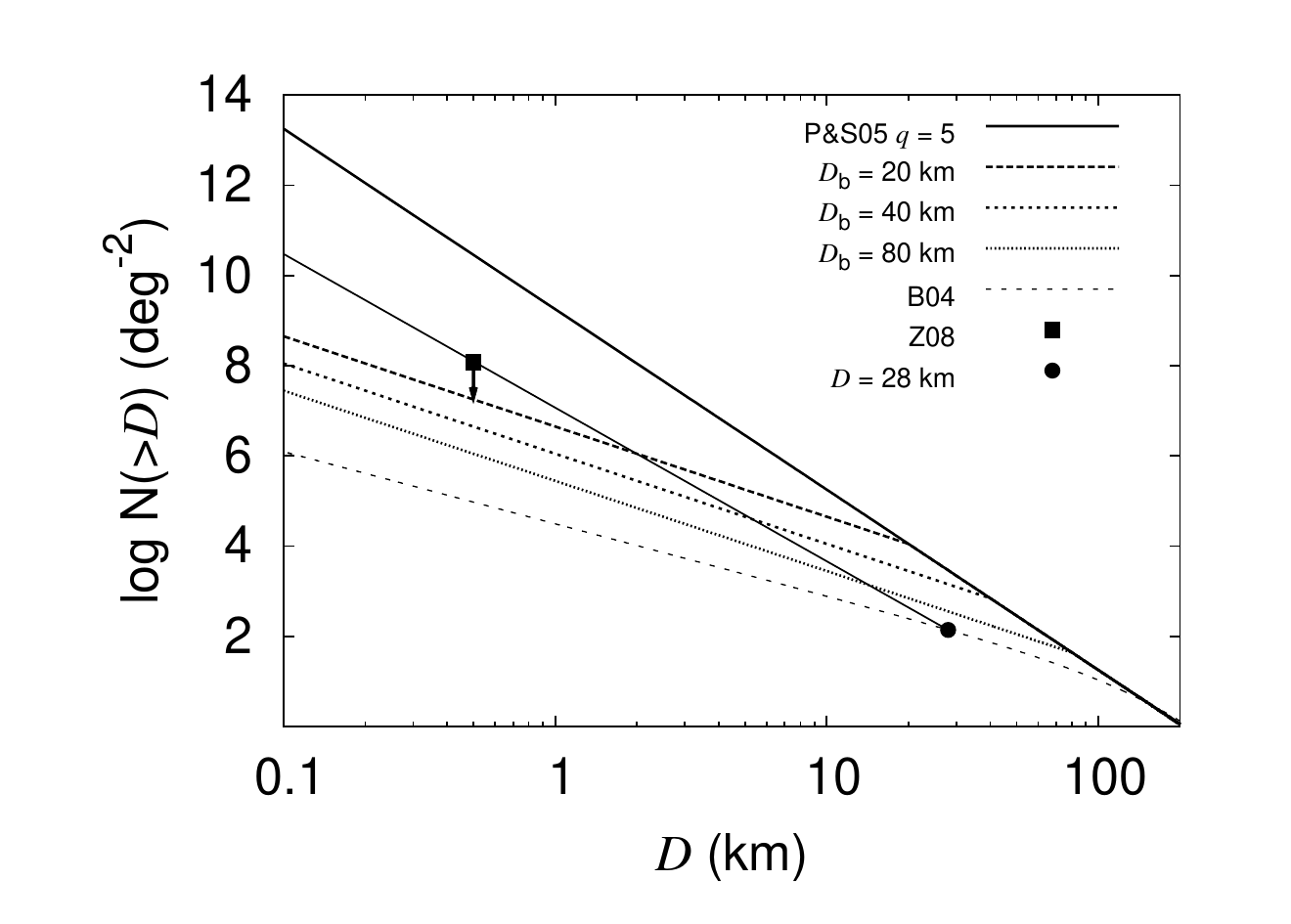}
\caption{The straight solid line (P\&S05) has $q=5$. The slope after the break diameter has $q=3$. Double power law from Bernstein has $\alpha1=0.88$ and $\alpha2=0.32$ (B04). TAOS (Z08) upper limit was anchored at 28~km on the B04 model and set an upper limit with $q<4.6$ at 0.5~km. The arrow indicates an upper limit.}
\label{fig:models}
\end{center}
\end {figure}

One can estimate the number of expected events as a combination the total solid angle,
detection efficiency and differential size distribution. It follows that: 
\begin{eqnarray}
\label{eq:events}
{\cal N}_{\rm event}  & \simeq  & \sum^{n}_{k} \sum_{j_k}^{n_{j_k}} v(\phi_{j_k},r)  \,  \, \xi_{j_k}   \, \frac{1}{r^{2}}\int_{0}^{\infty} \frac{dN}{dD} \times \, \nonumber \\
& & H\left(\lambda,r,D, \theta_{k} \right) \,  \epsilon (\lambda, D,\theta_{k}, S_{k}) \, dD  \,\, ,
\end{eqnarray} 
where $n$ is total number of guide stars, $j_k$ is the $j$th exposure for guide star $k$, $v(\phi_{j_k},r)$ is the apparent motion of the TNO
for guide star $k$ and exposure $j_k$, $\xi_{j_k}$ is the exposure time, 
$dN/dD$ is the differential size distribution,    
$H(\lambda, r, D, \theta_k)$ is the effective diameter,   
and $\epsilon (\lambda, D,\theta_{k}, S_{k})$ is the detection efficiency for PS1. 
The apparent motion $v(\phi_{j_k},r)$ and effective diameter $H_{k}(\lambda, r, D, \theta_k)$ can be calculated 
with $\phi=0^{\circ}$ (PS1 will observe mainly close to opposition) for all stars and all pointings, $r=43$~AU and $\lambda$'s corresponding to the particular
filters. The total number of exposures $n_{j_{k}}$  
for each star $k$ will be {\em reduced} according to the time that PS1 is pointing within the ecliptic. 
 If we assume that 
TNOs are distributed uniformly within $|\beta| \leq 6^{\circ}$ along ecliptic, the total time PS1 will  
spend near ecliptic is  then $\sin(\beta) \, \tau \simeq 0.10  \, \tau$, where $\tau$ is the total 
time of observation in seconds for a three-year lifetime.  The detection efficiency for PS1 has been investigated as a function of $\theta_k$ and $S_k$ in $\S$\ref{sec:det}. 
The last unknown in the event rate calculation is the model dependent differential size distribution. For this, we adapted three differential size distributions.
First, we used  a double power-law  distribution described in \citet{bernstein2004} (hereafter B04) with bright-end slope $\alpha_1=0.88$ and faint-end slope $\alpha_2=0.32$.  
The conversion from R-magnitude to size of the TNO was done by using 4\% albedo and $\phi \sim 0^{\circ}$ \citep{russell1916}.
The cumulative size distribution is shown as B04 line in \fig{fig:models}. 
Second, we use the models by \citet{pan2005} (hereafter P\&S05) with three different break diameters at 20, 40 and 80~km and $q=3$ also shown in \fig{fig:models}. The third model is from the upper limit set by the TAOS project  \citep{kiwi2008} (hereafter Z08). Anchored at 28~km, TAOS collaboration reported an upper limit on the cumulative surface density for objects larger than 0.5~km shown as solid circle (anchored point) and solid square (upper limit) also in \fig{fig:models}.  Other models, such as \citet{kenyon2004}, predicted the cumulative size distributions for sub-kilometer objects to be  within the ranges of the three models mentioned above and hence are not shown here.

\begin{figure}[]
\begin{center}
\includegraphics[width=90mm]{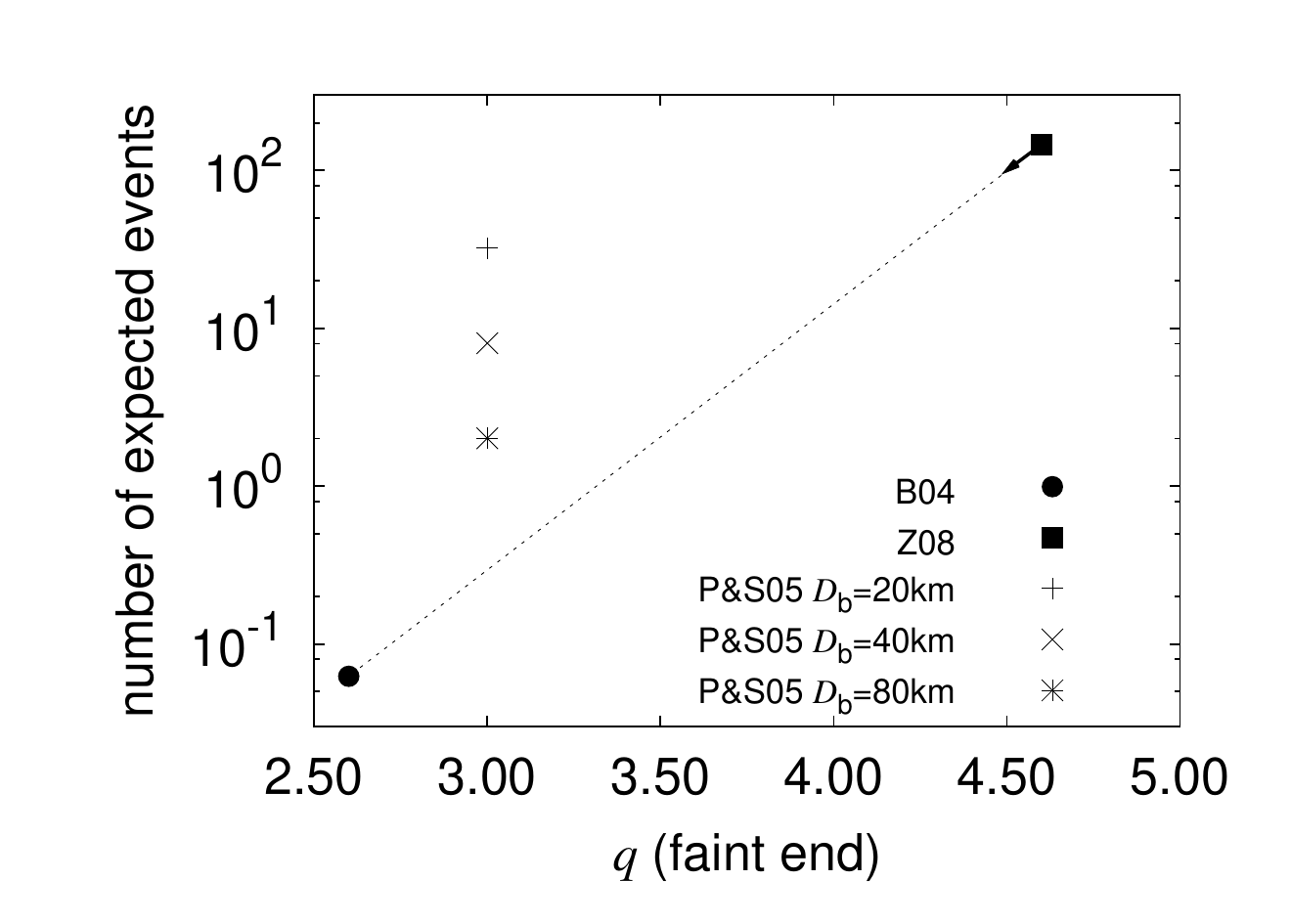}
\caption{Total number of expected events in the  PS1 three-year lifetime, assuming all 60 guide stars are 0.02~mas and SNR=100, for different models. The arrow between Z08 and B04 indicates that the number of expected events will move along the dash-line as the upper limit (i.e. $q$) becomes smaller (as stronger constraints apply to the upper limit). Number of expected events from Z08 gives the most optimistic value of $\sim 100$. The number of expected events from P\&S05's models range from a few to a few tens. 
For B04, we expect less than one event in the PS1 three-year lifetime.}
\label{fig:rate1}
\end{center}
\end {figure}

We next calculate the number of events for three models assuming 300 clear nights per year, 
eight hours of observation per night and 60 guide stars with 0.02~mas and SNR=100 for PS1 three-year lifetime as 
show in \fig{fig:rate1}. From the TAOS upper limit,  the most optimistic prediction for the number of events is $\sim$100. The P\&S05 model with different break diameter have number of events ranging from a few to a few tens.  For B04, we expect less than one event in PS1's three year life-time.

\section{Guide Stars Catalog}
\label{sec:catalog}
\begin{figure}[]
\begin{center}
\begin{tabular}{c}
\resizebox{90mm}{!}{\includegraphics{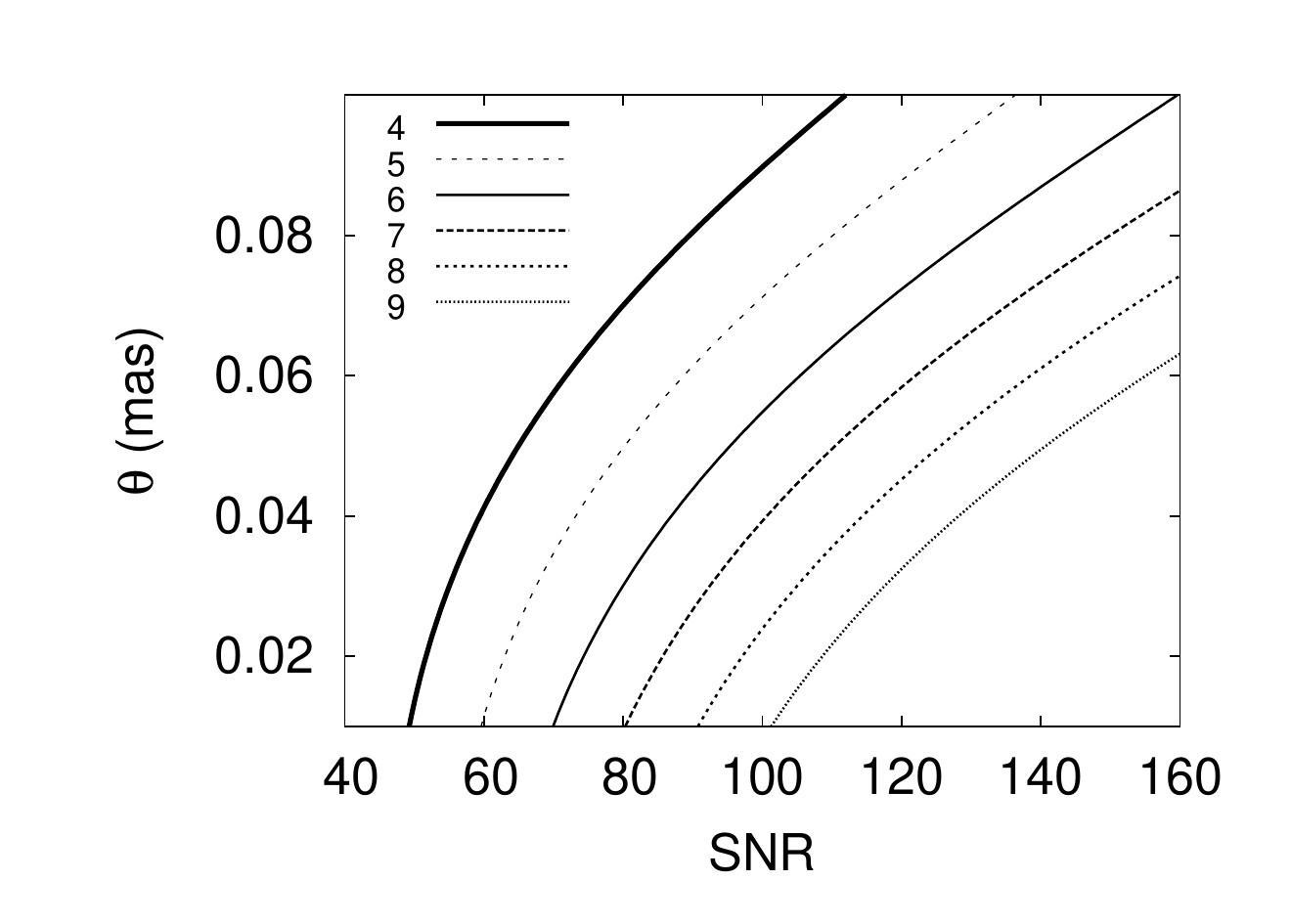}} 
\end{tabular}
\caption{Total number of expected events in the PS1 three-year lifetime. Given SNR, $\ts$ and   Pan \& Sari's model with 40~km break diameter with $q=3$, assuming all 60 guide stars have the  same angular sizes and SNRs.}
\label{fig:rate2}
\end{center}
\end {figure}

As we have shown in the previous section, the detection efficiency is higher for smaller stellar 
angular size and higher SNR. Therefore, it is crucial to select guide stars with the smallest possible 
angular size yet with good SNR in order to maximize the detection rates. 
Nevertheless, it is not clear whether a star with SNR=120, $\theta_{\star}$=0.07~mas is a better choice than another star with SNR=100 and $\theta_{\star}$=0.03~mas. To answer this question  we chose a size distribution  from models, such as 
P\&S05 with break diameter $D_{\rm b}=40$~km, and  calculated the number of expected events for a given star. 
In \fig{fig:rate2}, we show the total number of expected events per star in PS1's three-year lifetime as a function of  angular size and SNR. 
Using this relation, we can now estimate the event rates for all matched stars (i.e., candidate guide stars) from Tycho2 and 2MASS catalogs and then select stars that give the highest event rates as our guide stars. 

\begin{figure}[]
\begin{center}
\begin{tabular}{cc}
\hspace{-.8cm}
\resizebox{84mm}{!}{\includegraphics{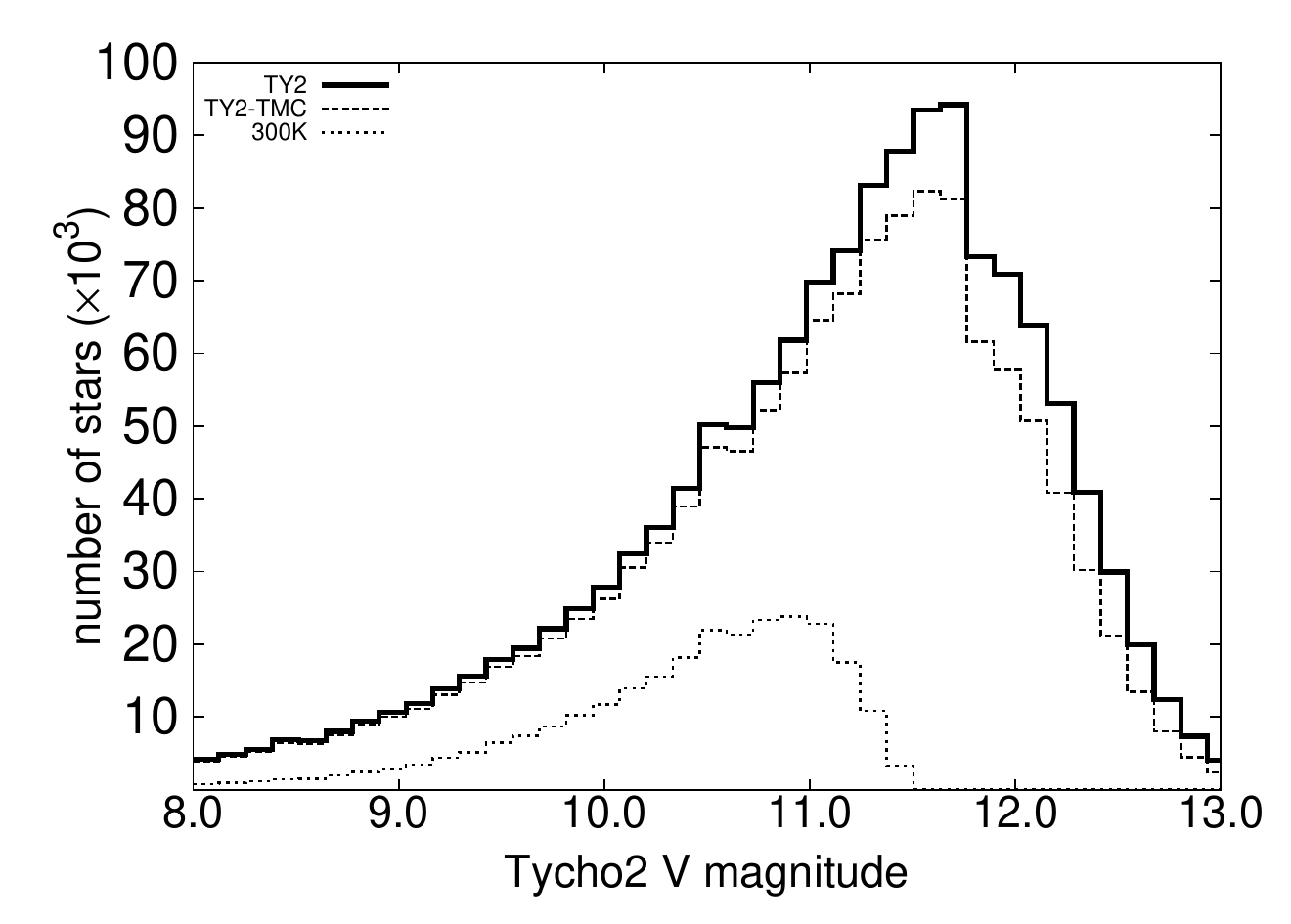}} &
\hspace{-1.0cm}
\resizebox{90mm}{!}{\includegraphics{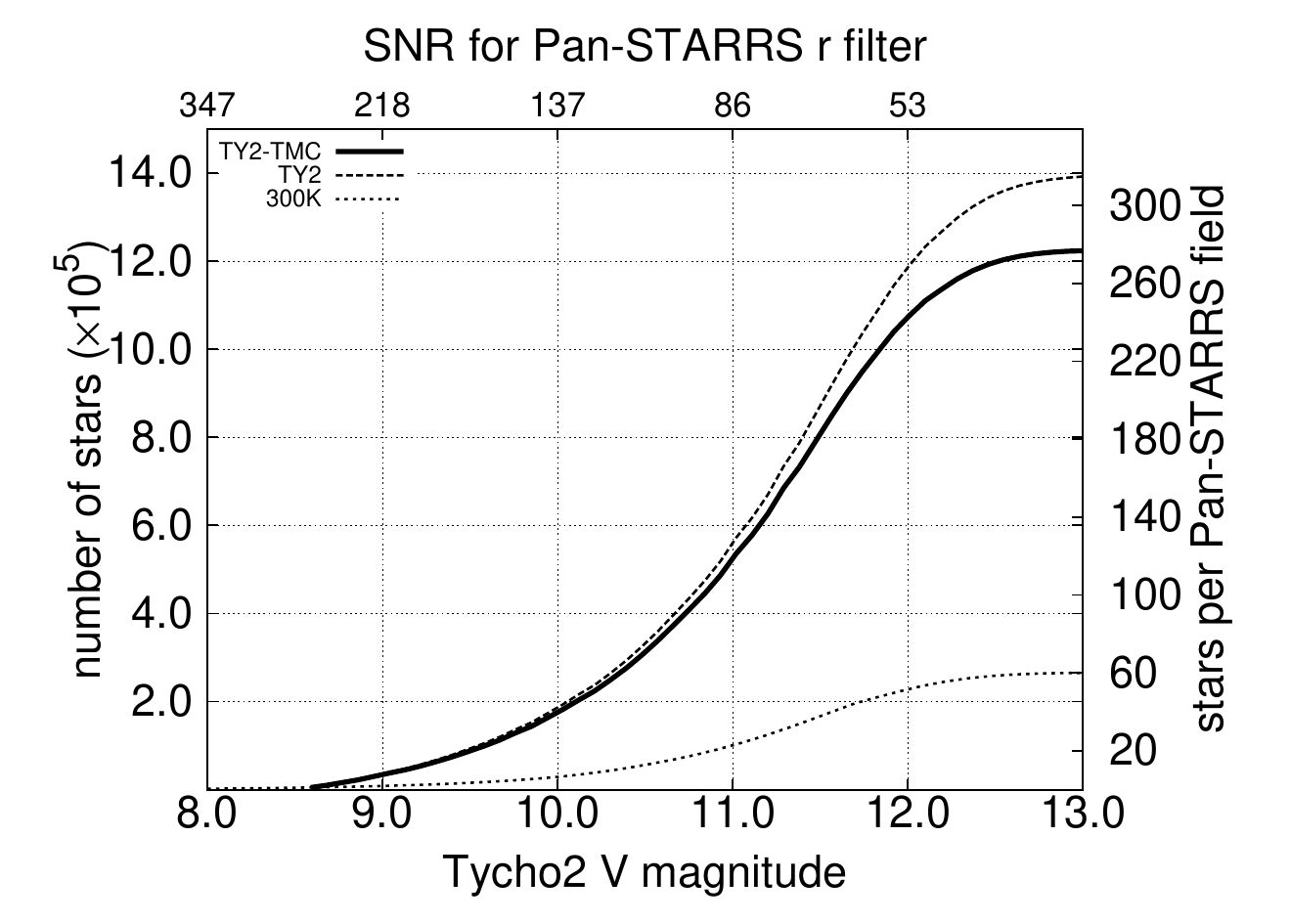}} \\
\end{tabular}
\caption{The left panel: Histograms of all stars visible to PS1 $3\pi$ sky in Tycho2 catalog with $m_{\rm V}<13$ mag (TY2), cross-matched stars between Tycho2 and 2MASS (TY2-TMC) and the top 300 thousands cross-matched stars (300K) that give the highest event rates based on their predicted angular sizes, SNRs and P\&S05 model.  The peak distribution is near $m_{\rm V} \sim 10.8$ mag which has predicted SNR $\sim$ 100. The right panel: Cumulative distribution. On average, there are 180 stars with $m_{\rm V}< 11.5$ mag for each PS1 7 deg$^{2}$ field; hence we can have multiple choices of guide stars ranked by their predicted event rates.}
\label{fig:vmag}
\end{center}
\end{figure}

In \fig{fig:vmag}, we show the distribution of $m_{\rm V}<13$ mag for all Tycho2 stars visible to PS1 $3\pi$ sky along with all matched stars between Tycho2 and 2MASS catalogs. For this work, we removed all matched stars with $m_{\rm V}<8.5$ mag due to possible saturation. As it can be seen, most of  the stars with $m_{\rm V} < 12$ mag were matched. On average, there are about 180 candidates with $m_{\rm V}< 11.5$ mag for every PS1 7 deg$^{2}$ field. 
Furthermore,  \fig{fig:rate2} shows the number distribution of matched  stars as a function of SNR and $\ts$. The distribution of SNR peaks around 80 and 
the angular size peaks around 0.02~mas.

 Considering that only 60 guide stars are needed for each target field, we  selected  $\sim 3\times 10^{5}$ guide stars from a total of $\sim1.27\times10^{6}$ candidates \footnote{For 3 $\pi$ sky and PS1 7 deg$^2$ field of view there are about 5000 unique target fields. }$^,$\footnote {We have compiled a complete list for all matched stars with $m_v<13$ mag. For the complete list see http://timemachine.iic.harvard.edu/andrew/ps.matched.all.uniq}. The dotted line in \fig{fig:vmag} is the distribution of the top  $3\times 10^{5}$ stars that were selected for the highest event rates. It is clear that most of these stars are also bright ($m_{\rm V}< 11.2$ mag). In reality, 
 even if the guide stars were selected based solely on their SNR, only a few stars per field would have not been 
 selected from our selection based on the event rates. For each PS1 target field, we provide multiple choices of guide stars ranked by the estimated event rates. Multiple choices will ensure freedom of selecting guide stars away from gaps in CCD arrays or dead pixels that would develop over time.

\section{Engineering Data}
\label{sec:engdata}
A set of video mode engineering data ($\sim 166$ stars-hours) was acquired in fall of 2008. The guide stars used in  these images were randomly picked bright stars as opposed to being selected 
based on the  optimized event rates  as we described in previous section. 
Each FITS \citep{fits} cube  in the data set contains a few thousand  $100\times50$~pixels images with one bright guide star. For the purpose of testing, these images were  taken with 10~Hz instead of 30~Hz sampling.  Nevertheless, we calculated all the related threshold values for 10~Hz sampling and tried to look for possible events in these engineering data. 

All  images in the FITS cube were processed with \emph{SExtractor} \citep{sextractor} to extract  lightcurves for every  guide star. Lightcurves for stars on the edges were excluded,  and the first and last 100 data points in the lightcurves were removed because no stars were in these images due to shutter operations.  
 
\begin{figure*}[]
\begin{center}
\begin{tabular}{cc}
\hspace{-.8cm}
\resizebox{90mm}{!}{\includegraphics{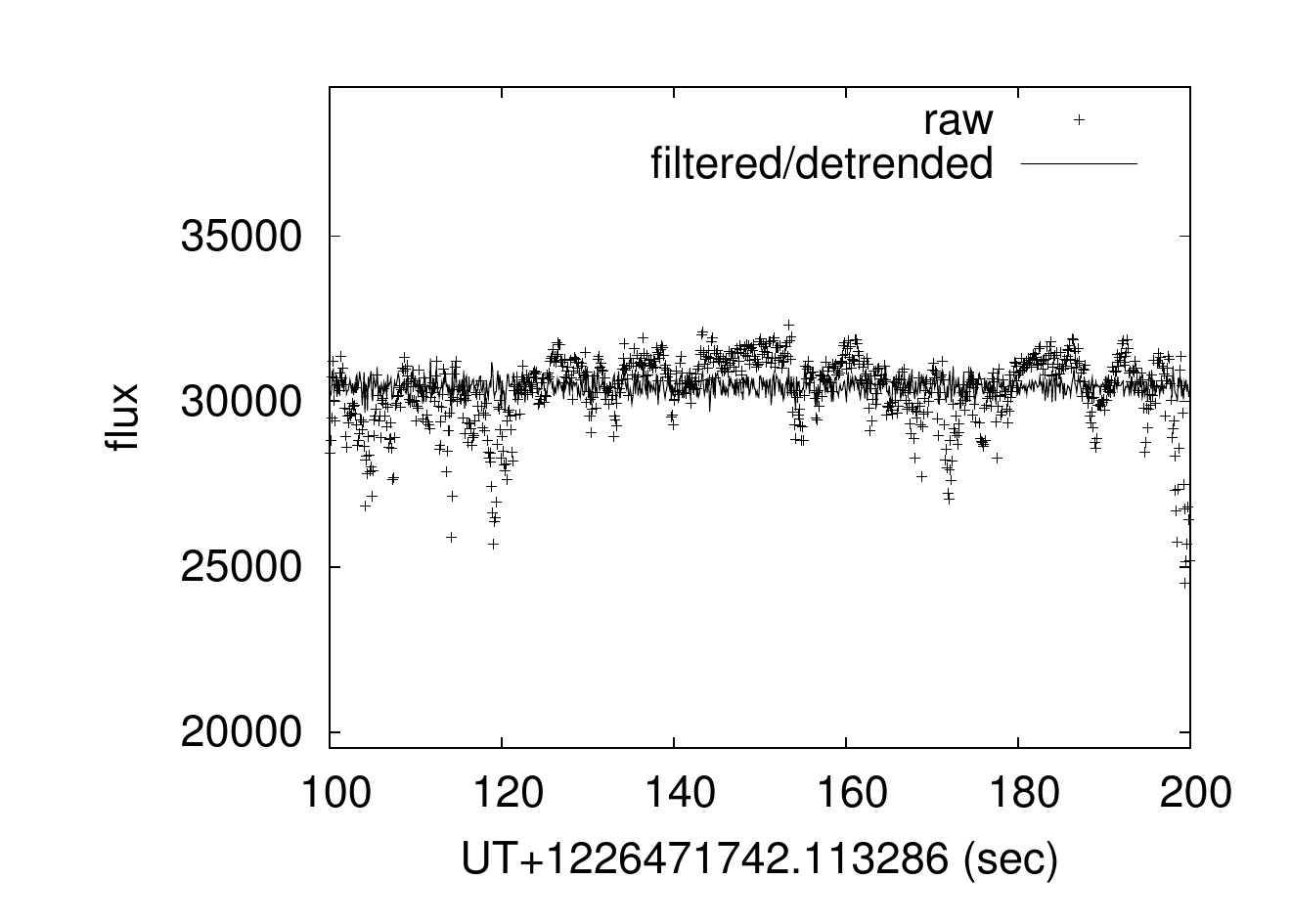}} &
\hspace{-1.2cm}
\resizebox{90mm}{!}{\includegraphics{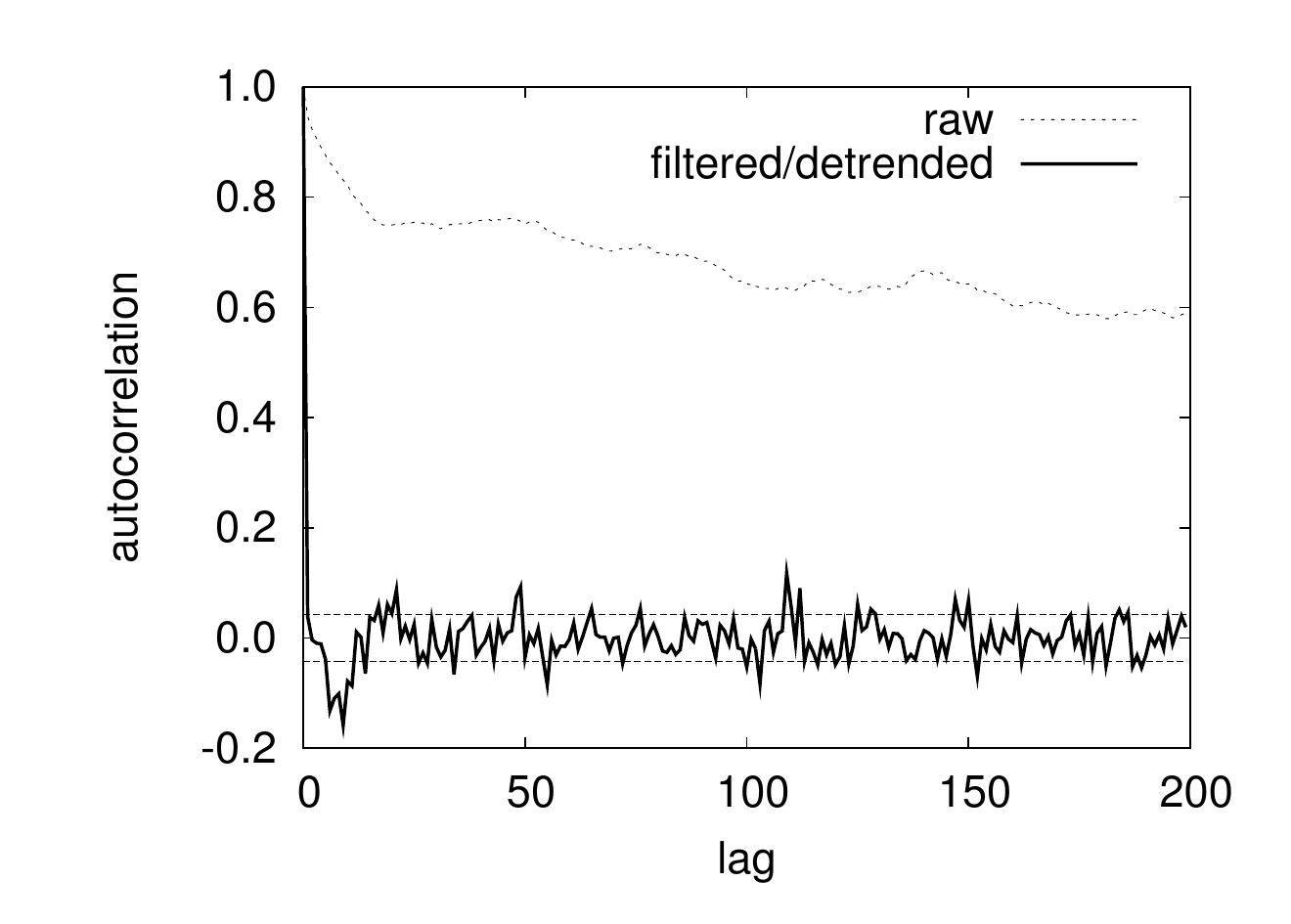}} \\
\hspace{-.8cm}
\resizebox{90mm}{!}{\includegraphics{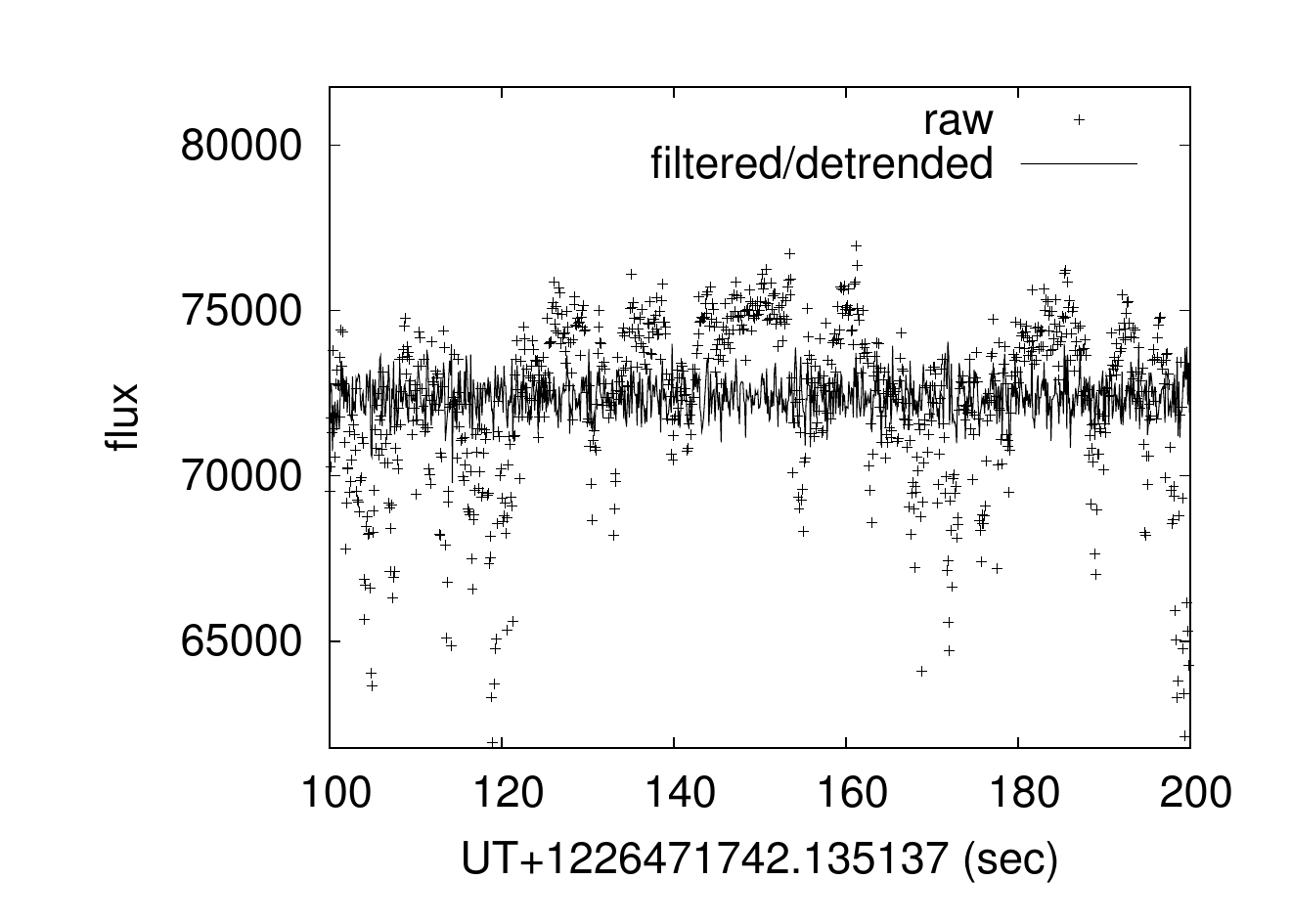}} &
\hspace{-1.2cm}
\resizebox{90mm}{!}{\includegraphics{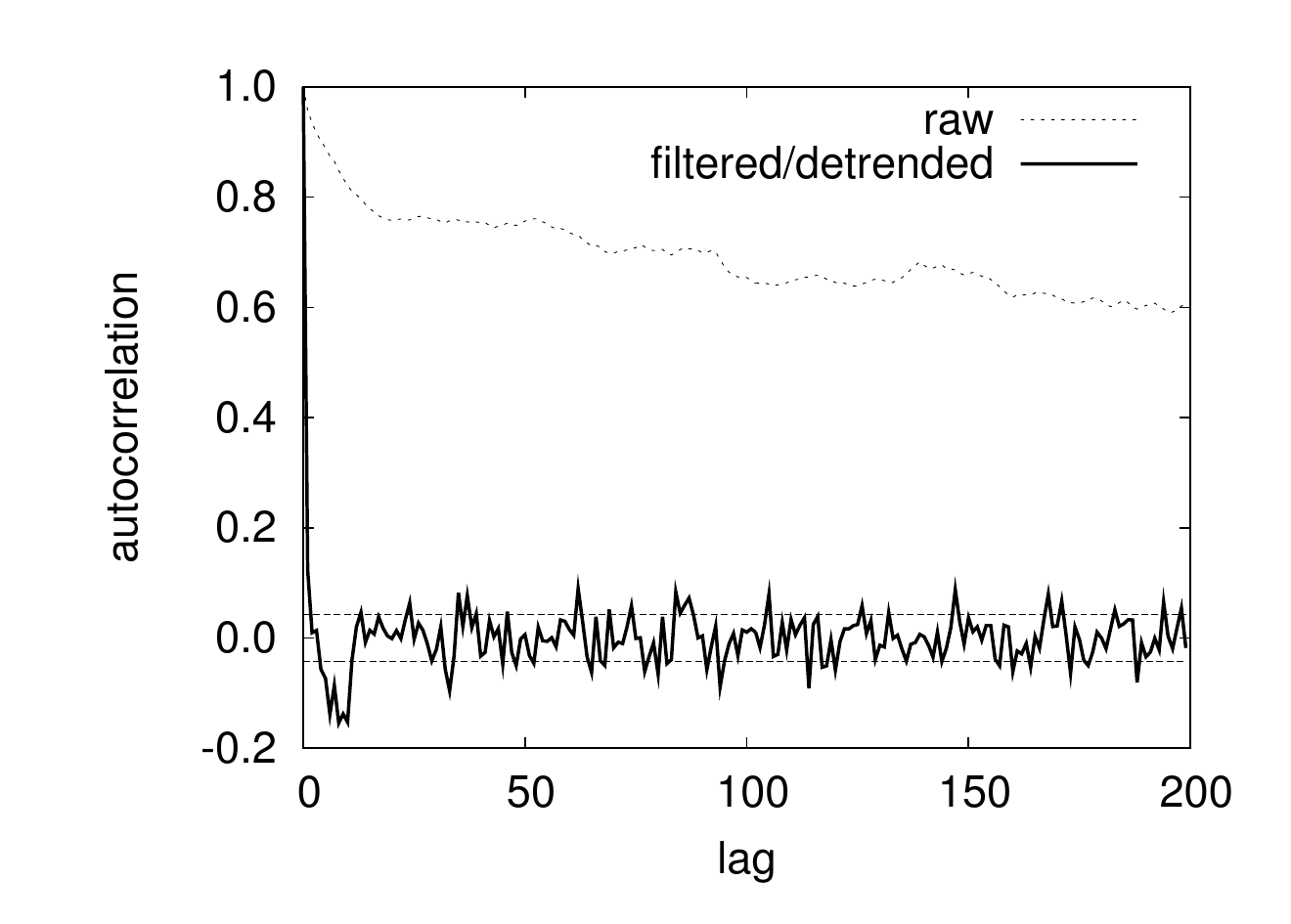}} \\
\end{tabular}
\caption{These two lightcurves and their autocorrelations  are from the same video mode batch. Left panel: The crosses  are the raw lightcurves  and the solid lines are the high-pass filtered and de-trended ones. High-pass filtering was done by removing a running mean of neighboring 21 data points from the lightcurves.  Right panel: The autocorrelations of the lightcurves before and after processing. The dashed lines are the  two-sigma range for autocorrelation of white noise \citep{box1970}. As it can be seen, high-pass filtering and de-tredning improved the SNR of the lightcurves and also removed the correlation from the lightcurves. }
\label{fig:eng}
\end{center}
\end{figure*}

 We  applied high-pass filter to the lightcurves  to remove the low frequency components (calculated by a running mean from 21 data points)  and then  de-trending to remove the common features existing among the lightcurves taken at the same time   \citep{kim2009}. \fig{fig:eng} shows two examples of lightcurves along with their autocorrelations before and after the lightcurves were high-pass filtered and de-trended.  The raw lightcurves showed considerable fluctuations due to either changing sky transparency or image motion, resulting in significantly lower measured SNR than predicted SNR from \eq{equ:snr}.  Measured SNR here is the median value divided by the standard deviation of the lightcurve.  Also, strong autocorrelations were present in the raw lightcurves, which can compromise the detection technique. Nevertheless, the filtered and de-trended lightcurves show  improved SNR and   
low  autocorrelations. Dashed lines in the right panels are the two-sigma range for autocorrelation of white noise: $(z-l)/z \,(z+2)$, where $z$ is the number of data points in the lightcurve and $l$ is the lag in the autocorrelation \citep{box1970}.

\begin{figure}[]
\begin{center}
\begin{tabular}{c}
\resizebox{90mm}{!}{\includegraphics{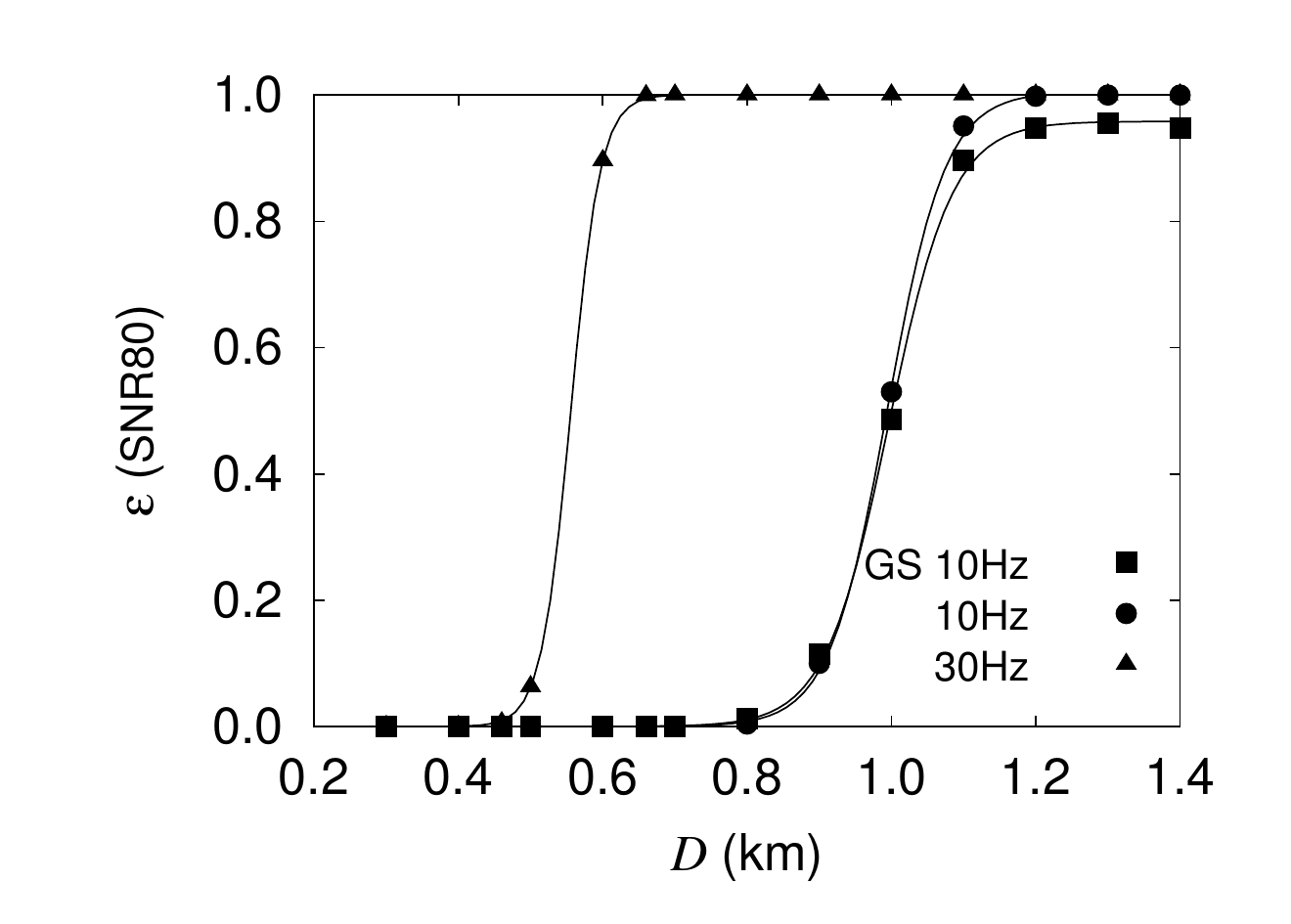}} \\
\end{tabular}
\caption{The detection efficiency of a guide star in the 10~Hz engineering data (solid square) with $\ts \sim 0.04$~mas, SNR $\sim$ 76.3. Events with various sizes were injected in the lightcurve to check the detection efficiency. Compared the detection efficiency of 30~Hz, it is clear that higher  sampling rate has better detection efficiency. Meanwhile, the efficiency from the 10~Hz simulation (solid circle) for similar SNR and $\ts$ was slightly better than the video mode lightcurve as expected. Most importantly, it also confirms that our detection algorithm works for PS1 lightcurve and can be used to search for TNO occultations.}
\label{fig:vmeff}
\end{center}
\end{figure}

Finally  we searched for possible events using the detection algorithm described in \sect{sec:det}. 
 To determine if the performance of detection method on the engineering data is consistent with the simulations. One of the guide star  ($\alpha=333.4820917^{\circ}$,  $\delta=1.2981389^{\circ}$)   lightcurves with $\ts \sim 0.04\,$mas and $m_{\rm V}=11.12$  mag and measured SNR $=76.3$ was selected and events with various sizes were injected. The detection method was then used to recover the injected events and therefore
 the efficiency (see  \fig{fig:vmeff}). Compared with the efficiency of 30~Hz for the similar SNR and $\ts$, one can see the detection efficiency of 30~Hz is better  than that of 10~Hz. This also confirms that higher sampling rate has better detection efficiency. Meanwhile, the efficiency from the simulation for similar SNR and $\ts$ was slightly better than the video mode lightcurve as expected.

\begin{figure}[]
\begin{center}
\begin{tabular}{cc}
\hspace{-.8cm}
\resizebox{90mm}{!}{\includegraphics{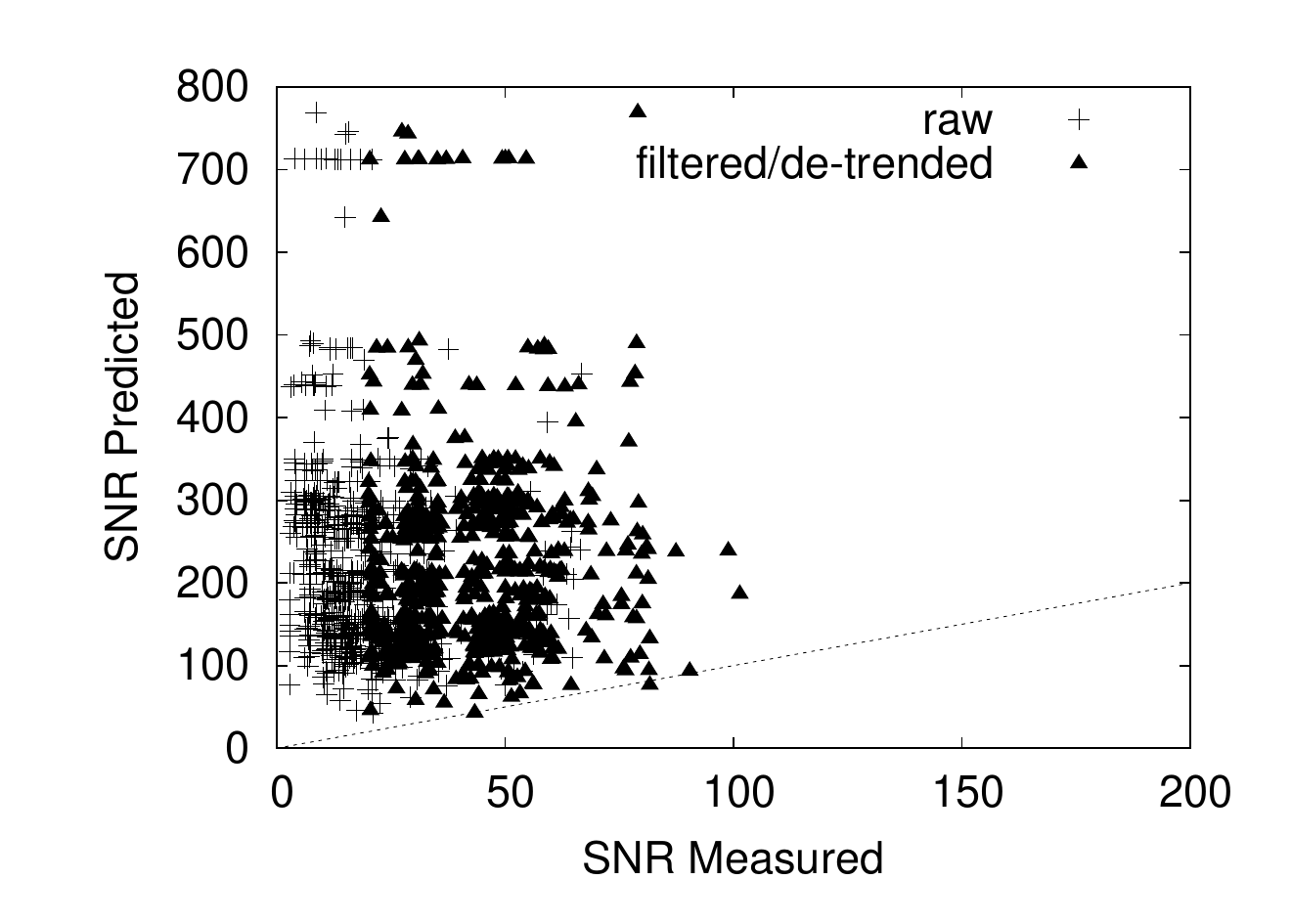}} &
\hspace{-1.0cm}
\resizebox{90mm}{!}{\includegraphics{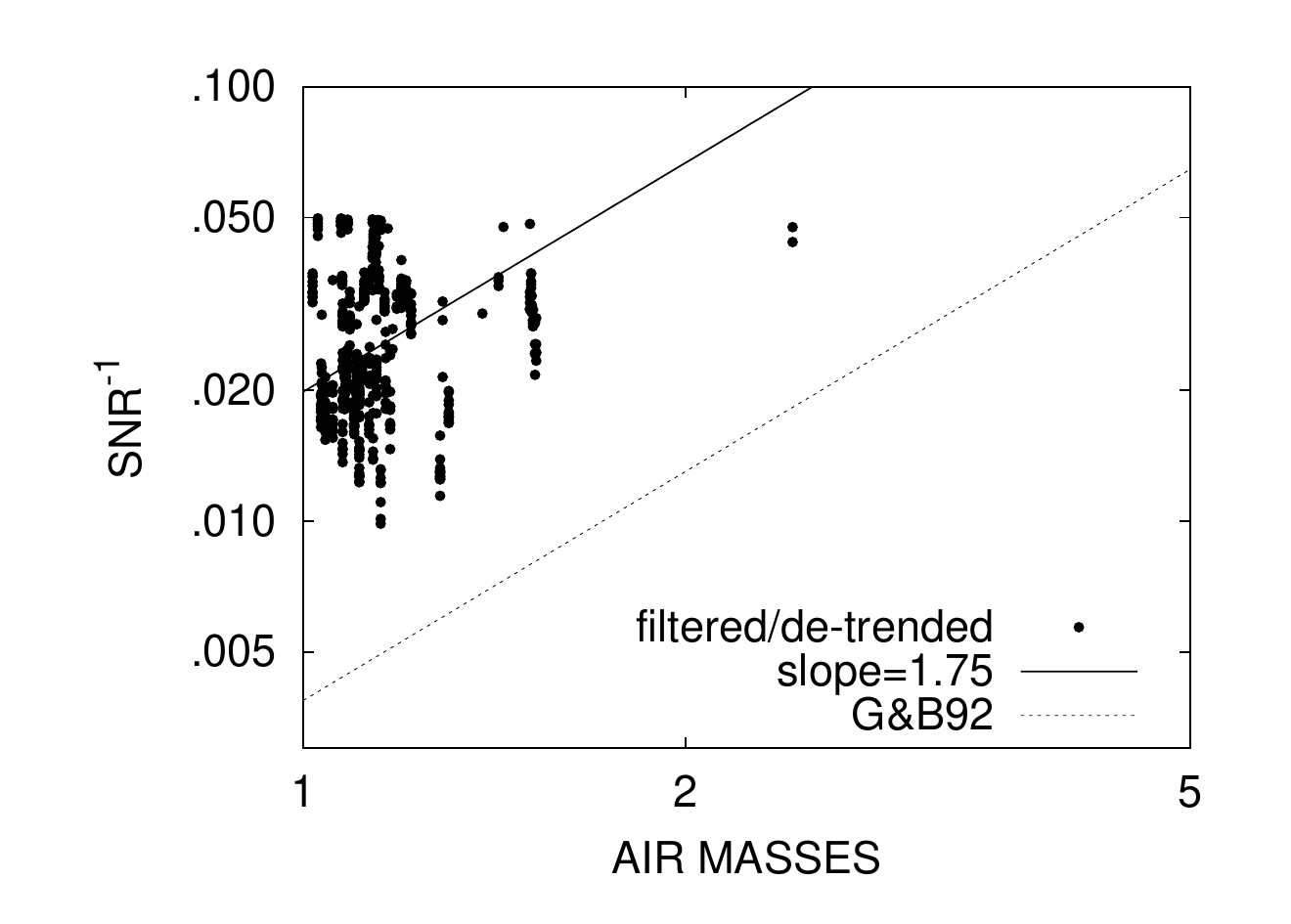}} \\
\end{tabular}
\caption{Left panel: The predicted SNR versus the measured SNR before and after filtering and de-trending. The dotted line has slope $=1$ to guide the eye. Even after filtering and de-trending, most of the measured SNRs were worse than the predicted ones. Right panel: The air masses versus the SNR$^{-1}$.  G\&B92 is the  scintillation noise limit described by \cite{gilliland1992}. The solid line is a fitting with fixed slope $=1.75$ of all measured SNR. The significant discrepancy suggests that other sources of noise, such as image motion, defocusing and unknown electronic noise,  also contributed to the measured SNR.}
\label{fig:vmsnr}
\end{center}
\end{figure}

To further evaluate the quality of the lightcurves, we compared the predicted SNR with the SNR of lightcurves before and after filtering and de-trending
(see left panel of \fig{fig:vmsnr}). As it can be seen in the plot, the measured SNR is never as good as the predicted ones.  In fact, the SNR of the lightcurves of bright guide stars 
never exceed $\sim200$, which suggests that the SNR is  limited by scintillation noise \citep{young1967, gilliland1992}.
The scintillation noise is estimated by Equation 11 in \cite{gilliland1992}:
 \begin{equation}
\label{equ:scin}
{\rm SNR}^{-1}=0.09A^{-2/3}(\chi)^{1.75}\exp(-h/h_{0})(2 \Delta t)^{-1/2},
\end{equation}
where $A$ is the diameter of the telescope in centimeters, $\chi$ is the air mass, $h=3000$~m is the altitude  at Haleakala,  $h_{\rm 0}=8000$~m is the scale height and $\Delta t$ is the exposure time in seconds.  The scintillation limit based on PS1 parameters are shown as dotted line in the right panel of \fig{fig:vmsnr}.  
The  engineering data has  SNR $<200$ which is  consistent with the scintillation noise with PS1 parameters. 
The immediate  impact of the reduction in the measured SNRs is the number of expected events. 

In \fig{fig:rate1}, we estimated the event rates from the distribution peak of predicted SNR(=100) and $\ts=0.02$~mas of the guide stars candidates. 
If the distribution peak of measured SNRs is reduced from 100 to 50,  
then the number of expected events will be halved (estimated from SNR = 50, $\ts$ = 0.02~mas in \fig{fig:rate2}). Nevertheless, given the wide range of expected  events depending on the models and other assumptions, we are still able to impose constraints and eliminate models.

We realized that these engineering data might not be taken from the best sky condition with optimal optical and mechanical performances. Significant discrepancy can be seen between the measured SNR and scintillation noise limit (\fig{fig:vmsnr}). Possible sources of the noise are image motion, defocusing and other unknown electronic noise. We expect the quality of the video mode images will  be improved in the future, and  better SNRs can be achieved as  of optical and mechanical problems are solved. In the end, no event in the engineering data set passed the one false positive threshold as was expected from our event rate calculations.

\clearpage

\clearpage

\begin{figure}[]
\begin{center}
\begin{tabular}{c}
\resizebox{90mm}{!}{\includegraphics{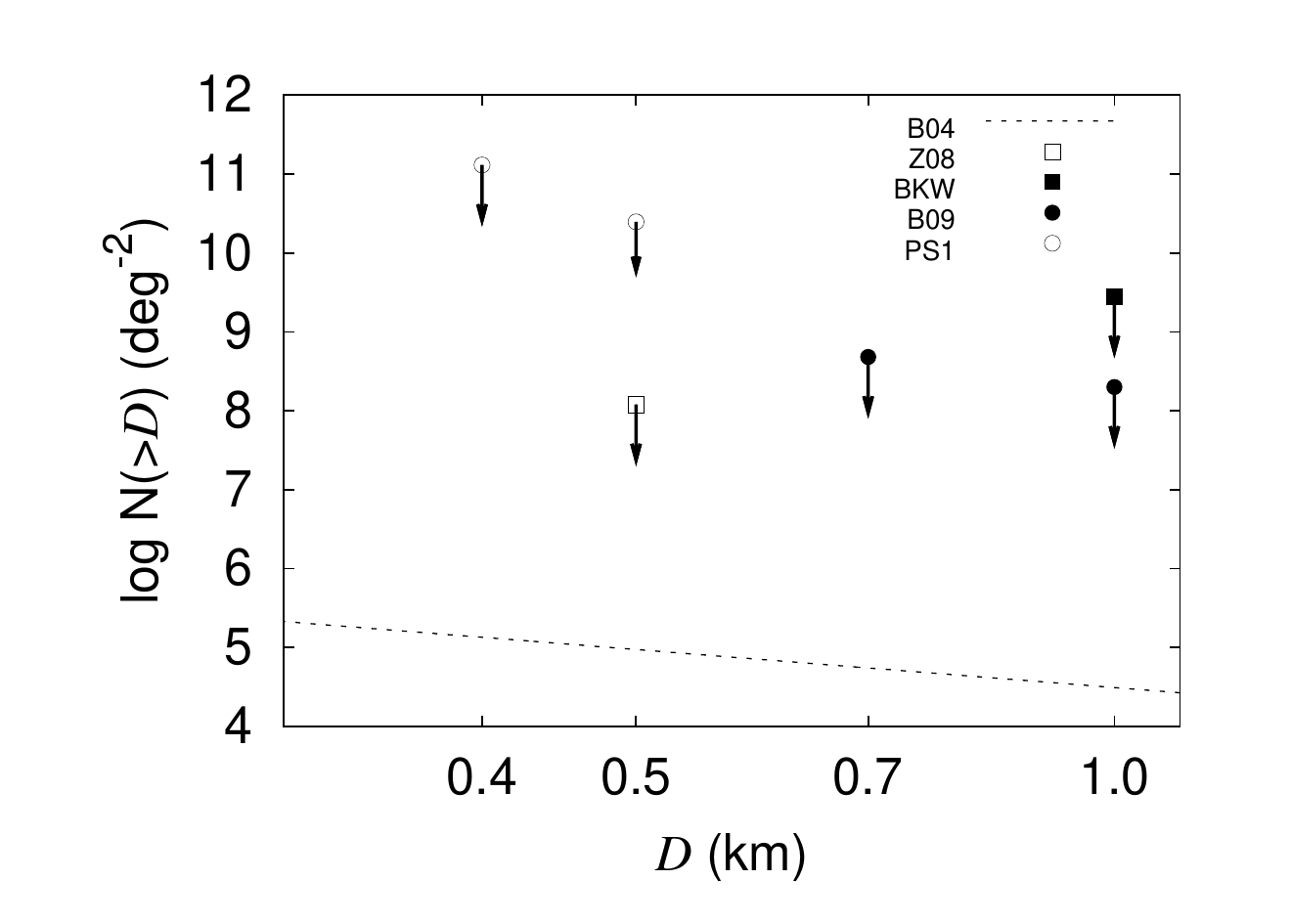}} 
\end{tabular}
\caption{The upper limit sets by PS1 22 star-hours engineering data at 95\% confidence limit along with upper limits by TAOS, BKW \citep{bickerton2008} and B09 \citep{Bianco2009}.}
\label{fig:upper}
\end{center}
\end {figure}

\section{Result and Discussion}
\label{sec:result}
We have made a pre-survey study of using lightcurves from guide star video mode images to search 
for occultation by TNOs near 43~AU. Simulations were made to calculate the null hypothesis distribution (\fig{fig:nulldist1}) and  detection efficiency for various angular sizes and SNRs. Under 30~Hz sampling, PS1 can detect objects as small as $\sim400$ m at 43~AU (\fig{fig:efficiency}). Using $(V-K)$ indices, we predicted the angular sizes of all matched stars between Tycho2 and 2MASS catalogs with $m_{\rm V}<13$ mag and $m_{\rm K}<16$ mag above the PS1 southern declination limit $\delta \ge-30^{\circ}$. The distribution of stellar  angular sizes peak around 0.02~mas in PS1 $3\pi$ sky (\fig{fig:vmk}). On average, there are about 
 180 stars with $m_{\rm V}<11.5$ mag for PS1 7 deg$^{2}$ field of view.  Given that we can choose only 60 
 guide stars for each PS1 target field,  we ranked our guide star candidates  by their number of expected 
 events based on their angular sizes, SNRs and model by P\&S05. As mentioned before, multiple choices 
 of guide star will ensure us the freedom of selecting guide stars away from gaps in the CCD array or dead 
 pixels that would develop over time. Based on the differential size distribution from different models and 
 the threshold  to have one false positive in the PS1 three-year lifetime, we estimated the number of expected 
 events could be somewhere from 1 to $\sim100$ (\fig{fig:rate1}).  The engineering data allowed us to 
 investigate the quality of the lightcurve and develop the detection pipeline for the upcoming real data. 
 We have established that the detection technique performs as well with the filtered engineering data. 
We also realize that the true SNR is limited by systematics. Some of the systematics will be removed 
as we move to real data stream, however some of the systematics will remain. For example, 
scintillation will limit the SNR to about 200. We have recalculated the event rates with the worst case 
scenario and  found that the event rates were compromised by a factor of four. Even with this pessimistic 
estimation the event rate that PS1 will find can allow us to place a constraint on the size distribution and 
hence the evolution of the TNOs. Using the available engineering data and detection efficiency at 0.4 and 0.5~km, 
we were able to derive an effective solid angle $\sim 2.3\times 10^{-11}$ and $\sim 1.2\times 10^{-10}$ deg$^{2}$ and set 
 the 95\% confidence upper limit on surface number density at
 $N(>0.4 \, \rm km)\sim 1.3\times10^{11}$  and 
 $N(>0.5 \, \rm km)\sim 2.47\times10^{10}$ deg$^{-2}$  (\fig{fig:upper}). 
In future work, for the upcoming real data, we will set the threshold based on  maximum true 
 positive to false positive ratio, which allows more candidate events for further investigations. Meanwhile, 
 the video mode lightcurves can also be used to search for objects in Sedna like orbits from 
 100 to 1000~AU. We will work on a new detection algorithm that is capable of searching for objects in this region.

\acknowledgements Work at NCU was supported by the grant NSC
96-2112-M-008-024-MY3. Work at the CfA was supported in part by the
NSF under grant AST-0501681 and by NASA under grant NNG04G113G. Work
at ASIAA was supported in part by the thematic research program
AS-88-TP-A02. And we also like to thanks IIC for hosting J-H Wang for his staying.

\bibliographystyle{apj}

\end{document}